\documentclass[preprint, 12pt, superscriptaddress,titlepage]{revtex4-2}

\usepackage{amsmath}
\usepackage{amssymb}
\usepackage{bm}
\usepackage[utf8]{inputenc}
\usepackage[english]{babel}
\usepackage{graphicx}
\usepackage{float}
\usepackage[section]{placeins}  % force floats into their section
\usepackage{booktabs}
\usepackage{multirow}
\usepackage{makecell}
\usepackage{afterpage}
\usepackage{textcomp}
\usepackage{xfrac}
\usepackage[unicode=true, colorlinks=true]{hyperref}
\begin{document}
\renewcommand{\abstractname}{}    % clear the "Abstract" title
\renewcommand{\figurename}{Fig.}
\makeatletter  \renewcommand\@biblabel[1]{#1.}  \makeatother
\newcommand{\parameters}{\boldsymbol{w}}

\title{Simulation-based inference of single-molecule force spectroscopy}
\bibliographystyle{naturemag}

\author{Lars Dingeldein}
\affiliation{Frankfurt Institute for Advanced Studies, 60438 Frankfurt am Main, Germany.}
\affiliation{Goethe University, 60438, Frankfurt am Main, Germany}
\author{Pilar Cossio}
\affiliation{Center for Computational Mathematics, Flatiron Institute, 10010 New York, United States}
\affiliation{Center for Computational Biology, Flatiron Institute, 10010 New York, United States}
\author{Roberto Covino}
\email[Corresponding author:]{covino@fias.uni-frankfurt.de}
\affiliation{Frankfurt Institute for Advanced Studies, 60438 Frankfurt am Main, Germany.}
\affiliation{International Max Planck Research School on Cellular Biophysics, 60438 Frankfurt am Main, Germany}

%  Main text – up to 3,500 words, excluding abstract, Methods, references and figure legends.
%    Abstract – up to 150 words, unreferenced. 
%    Display items – up to 6 items (figures and/or tables). 
%    Article should be divided as follows: 
%        Introduction (without heading) 
%        Results
%        Discussion
%        Online Methods.
%
%    Results and Methods should be divided by topical subheadings; the Discussion does not contain subheadings.
%    References – as a guideline, we typically recommend up to 50.

\begin{abstract}

%Single-molecule force spectroscopy (smFS) is a powerful approach to studying molecular self-organization. However, the interpretation of these experiments remains a challenge. 
%
%Coupling the molecule with the ever-present experimental device introduces artifacts. Performing statistical inference to learn molecular properties is challenging because even minimal models lead to intractable likelihoods. 
%
%e develop a computational framework that performs Bayesian inference to extract reduced quantitative models from smFS. Using synthetic data, we could systematically disentangle the measurement of hidden molecular properties from experimental artifacts. 
%
%We built on novel statistical methods called simulation-based inference (SBI). SBI enabled us to directly estimate the Bayesian posterior by encoding a mechanistic model into a simulator in combination with probabilistic deep learning.
%
%By integrating physical models with machine learning density estimation, we could perform accurate Bayesian inference for models with an intractable likelihood, paving the way for more realistic models so far ruled out due to their mathematical intractability. Our framework is general, transparent, easy to use, and broadly applicable to other types of biophysical experiments.

Single-molecule force spectroscopy (smFS) is a powerful approach to studying molecular self-organization. However, the coupling of the molecule with the ever-present experimental device introduces artifacts, that complicates the interpretation of these experiments. 
Performing statistical inference to learn hidden molecular properties is challenging because these measurements produce non-Markovian time series, and even minimal models lead to intractable likelihoods. 
To overcome these challenges, we developed a computational framework built on novel statistical methods called simulation-based inference (SBI). SBI enabled us to directly estimate the Bayesian posterior, and extract reduced quantitative models from smFS, by encoding a mechanistic model into a simulator in combination with probabilistic deep learning. Using synthetic data, we could systematically disentangle the measurement of hidden molecular properties from experimental artifacts. The integration of physical models with machine-learning density estimation is general, transparent, easy to use, and broadly applicable to other types of biophysical experiments.
\end{abstract}

\maketitle

\section{Introduction}

Single-molecule experiments provide an invaluable tool for understanding how molecules self-organize in cells and complex materials. These experiments quantify the dynamics of individual molecules, capturing their heterogeneity and stochasticity. They are instrumental in understanding molecular self-assembly phenomena, like folding, the process by which proteins, nucleic acids, and other polymers form well-defined 3D structures.

Single-molecule force spectroscopy (smFS) is a powerful approach to investigating the microscopic mechanisms of folding and other structural rearrangements \cite{petrosyan_single-molecule_2021}. It can reveal folded and unfolded states, short-lived intermediates, characterize the transition paths connecting metastable states and binding and unbinding events \cite{wang_templated_2022, kramm_dna_2020, pelz_subnanometre_2016, bustamante_optical_2021, petrosyan_unfolded_2021} 
%If it fits Michael Schlierf (maybe reviewer): https://pubs.acs.org/doi/full/10.1021/acs.nanolett.5b03956 
Typically, a globular biomolecule will mainly populate its folded state and only rarely unfold and refold again. In smFS, two handles are attached to the biomolecule and used to apply mechanical tension to it. This tension destabilizes the folded state and promotes unfolding. In experiments at constant force, the biomolecule is in quasi-equilibrium and repeatedly unfolds and refolds. By monitoring an order parameter, e.g., the molecule's extension, we could obtain a one-dimensional time series showing hopping between the folded (low extension) and unfolded (high extension) states. We could then estimate the populations and lifetimes of each state as a function of the applied force \cite{evans_dynamic_1997, hummer_kinetics_2003, dudko_beyond_2003, dudko_intrinsic_2006, hyeon_multiple_2012, Cossio2016}.

\begin{figure}[htp]
    \centering
    \includegraphics[width=0.9\textwidth]{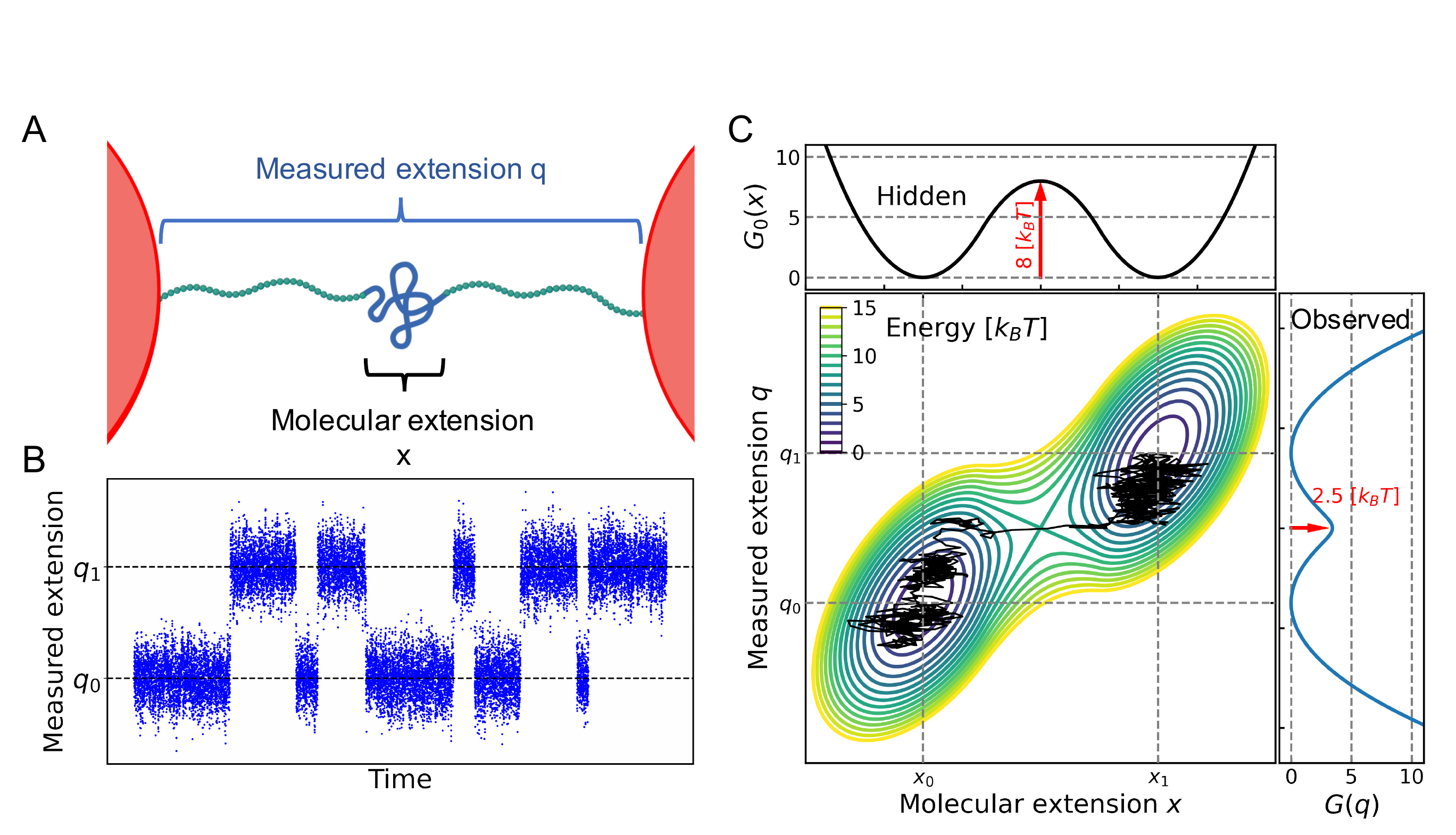}
    \caption[]{\textbf{Schematic modelling of a smFS experiment at constant tension}.(A) Schematic representation of a smFS experiment at constant force. The red spheres represent the mesoscopic beads that are used in optical tweezer experiments to apply force. The molecule of interest (blue) is attached via flexible polymer linkers (green) to the pulling device. The measured extension $q$ includes the length of the polymer linkers plus the extension of the molecule. (B) A time series $\boldsymbol{q}_t$ modelling an observation from a smFS experiment at constant force. The bi-stable trajectory shows rapid stochastic transitions between two states. (C) Example of the two-dimensional free energy surface $G(q,x)$. Isolines are drawn every 1 $k_{\mathrm{B}}T$. The black curve shows a representative trajectory transitioning between the states. The upper panel shows the molecular free energy profile $G_0(x)$, with a barrier height $\Delta G^{\ddagger} = 8~k_{\mathrm{B}}T$. The right panel shows the observed free energy $G(q)$---the potential of mean force along $q$---with a projected  barrier height of approximately $2.5~k_{\mathrm{B}}T$.}
    \label{fig: SMFE}
\end{figure}

However, the influence of the measuring apparatus---a pulling device attached to the small molecule via long flexible linkers (Fig. \ref{fig: SMFE}A)---significantly affects the measurements and complicates a quantitative interpretation of smFS. Ideally, we would directly monitor the dynamics of the molecular extension $x$ and measure a time series $\boldsymbol{x}_t$. In practice, we have only access to the measured extension $q$  (Fig. \ref{fig: SMFE}B), a combination of the molecular and linker extensions. The measuring apparatus is a mesoscopic object, much larger and slower than the molecule. The linkers are flexible and respond relatively slow to forces. Therefore, the time series of the measured extension $\boldsymbol{q}_t$ reports only indirectly on the molecular extension. 
Ignoring this effect leads to significant artifacts \cite{dudko_locating_2011,maitra_influence_2011, friddle_interpreting_2012,pierse_kinetics_2013,Makarov2014,hinczewski_mechanical_2013,Cossio2015,Cossio2018,covino_molecular_2019,satija_broad_2020,mondal_energy_2021}.  
Ingenious methods exist to disentangle measuring artifacts from measurements \cite{Stigler2011,turkcan_bayesian_2012,bryan_inferring_2020,bryan_inferring_2022} or to reduce the apparatus's distortion \cite{Yu2017,Woodside2006a}. Yet, these methods are often challenging to apply or lack generality. Some are only valid if the diffusion coefficient of the pulling device is as fast as the molecular one, which is in general not true.
Formulating a general and systematic framework to extract reduced quantitative models from smFS that recapitulate the molecular thermodynamic and kinetic properties is still an open challenge \cite{petrosyan_single-molecule_2021}.

Simulation-based inference (SBI) is a powerful technique to perform Bayesian inference to connect observations to mechanistic models \cite{cranmer_frontier_2020}. SBI is particularly suited for systems with an intractable likelihood, i.e., with no closed analytical form, and computationally expensive to evaluate \cite{lueckmann_benchmarking_2021}. The main idea is to encode into a simulator a parametric mechanistic model of an experimental observation. Given specific parameter values, the simulator produces synthetic data. Parameters that lead to synthetic data close to the original observation are the most plausible ones explaining it. Advances in density estimation due to neural networks and deep learning enabled a new generation of powerful SBI methods, which produce surrogate models of the likelihood or posterior using simulated data \cite{papamakarios_fast_2016}. SBI is a general approach \cite{zeraati_flexible_2022} and a growing field with broad applications ranging from particle physics \cite{brehmer_simulation-based_2021}, cosmology and astrophysics \cite{lemos_robust_2022, dax_real-time_2021}, to genomics \cite{bernstein_addressing_2021} and neuroscience \cite{lueckmann_flexible_2017, goncalves_training_2020}. 

Here, we develop a computational framework that performs Bayesian inference to build quantitative models from smFS experiments at constant force. We show how we can easily extract``hidden" molecular properties using only the measured time series $\boldsymbol{q}_t$. We overcome the intractable likelihood problem by using neural density estimation to directly estimate the Bayesian posterior. Our approach is general, conceptually transparent, easy to use, and broadly applicable to other types of biophysical experiments.

\begin{figure}[htp]
    \centering
    \includegraphics[width=0.9\textwidth]{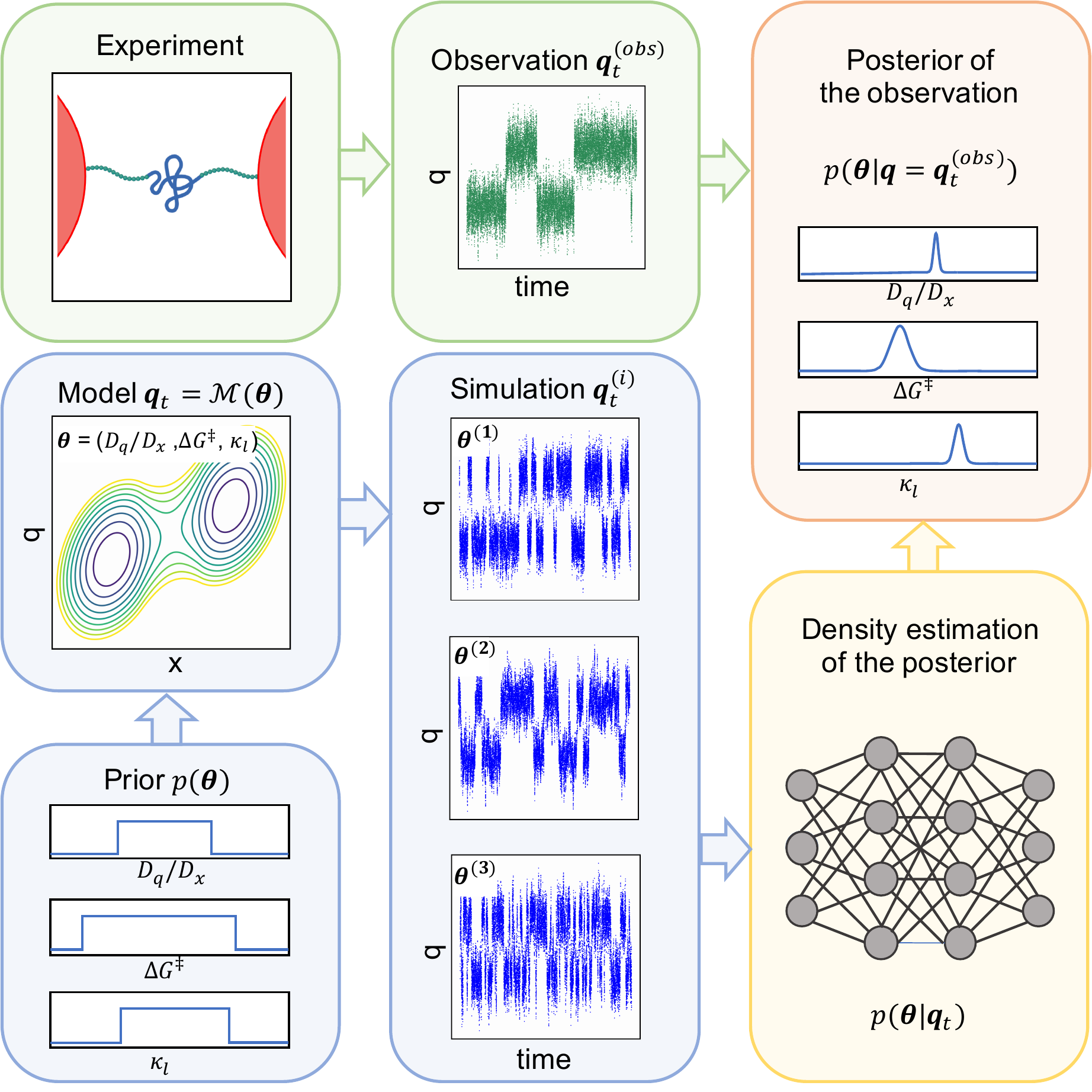}
    \caption[]{\textbf{Workflow of Simulation-Based Inference of smFS experiments}. Given an experimental observation $\boldsymbol{q}^{\mathrm{\mathrm{(obs)}}}_t$, we want to explain it with a parametric mechanistic model $\mathcal{M}(\boldsymbol{\theta})$ encoded in a simulator. The parameters $\boldsymbol{\theta}$ are drawn from the prior distribution $p(\boldsymbol{\theta}$). The simulator takes parameters drawn from the prior and generates synthetic observations $\boldsymbol{q}_t^{\mathrm{(i)}}$. The synthetic observations and corresponding parameters are used to train a conditional density estimator to approximate the posterior $p(\boldsymbol{\theta} | \boldsymbol{q}_t)$. Evaluating the posterior, $p(\boldsymbol{\theta} | \boldsymbol{q}_t=\boldsymbol{q}^{\mathrm{\mathrm{(obs)}}}_t)$, we obtain the most plausible parameters explaining a given observation.}
    \label{fig: simulation-based inference}
\end{figure}

\section{Results}
\subsection{Diffusive models of smFS with hidden degrees of freedom}

We aim to extract reduced quantitative models from smFS experiments based on a diffusive Brownian dynamics on a 2-dimensional free energy landscape \cite{Hummer2005}. The experiment will measure the 1-dimensional time series of the measured extension $\boldsymbol{q}_t = \{q_{t{\Delta \tau}}\}_{t=1}^{M}$, recorded with a lag-time $\Delta \tau$ and containing $M$ data points. The measured extension will indirectly report on the molecular extension, described by the time series $\boldsymbol{x}_t = \{x_{t{\Delta \tau}}\}_{t=1}^{M}$. Ideally, from $\boldsymbol{x}_t$ we could estimate a probability distribution $P(x)$ at a given force and then get the force-dependent free energy profile $G(x) = -k_{\mathrm{B}}T\log P(x)$, where $T$ is the absolute temperature and $k_{\mathrm{B}}$ is Boltzmann's constant. We could also estimate the diffusion coefficient $D_x$ from the fluctuations of $x$ \cite{Hummer2005}. In practice, all these quantities are hidden by the compounded dynamics of the measuring apparatus. 

We consider here a well-established minimal model to describe the joint dynamics of the molecule and apparatus, first introduced by Hummer and Szabo \cite{Hummer2010,Cossio2015,covino_molecular_2019}, where $\boldsymbol{q}_t$ and $\boldsymbol{x}_t$ are diffusive processes on the two-dimensional free energy surface $G(q,x) = G_0(x) + \frac{\kappa_l}{2}(x - q)^2$ (Fig. \ref{fig: SMFE}C). The molecule's extension $x$ diffuses with diffusion coefficient $D_x$ on the molecular free energy profile $G_0(x)$. The measured extension $q$ is coupled to $x$ by an harmonic linker term, with stiffness $\kappa_l$, and diffuses with $D_q$. Both parameters describe the pulling device's properties. Let us first consider the simple case where $G_0(x)$ is an ideal symmetric double-well, with a barrier of height $\Delta G^{\ddagger}$ separating folded and unfolded states. The challenge becomes to estimate the parameters $\boldsymbol{\theta} = \{\Delta G^{\ddagger}, D_q/D_x, \kappa_l\}$  by using only the measured extension $\boldsymbol{q}_t$. 
%This time series is the outcome of a projection of the full-dimensional system onto the measured extension and is non-Markovian. 
By naively estimating the free energy profile $G(q)$ from $\boldsymbol{q}_t$, we would obtain a significantly biased value of the free energy barrier (Fig. \ref{fig: SMFE}C).

Bayesian inference provides a general framework to estimate the "hidden" molecular parameters from the measured extension. The result of the inference is the posterior $p(\boldsymbol{\theta} | \boldsymbol{q}_t)$, a probability distribution that quantifies how much the parameter values are compatible with the observed trajectory $\boldsymbol{q}_t$. The posterior distribution is the outcome of Bayes' theorem
\begin{equation}\label{eq: bayes_theorem}
    p(\boldsymbol{\theta} | \boldsymbol{q}_t) = \frac{p(\boldsymbol{q}_t | \boldsymbol{\theta}) \cdot p(\boldsymbol{\theta})}{\int p(\boldsymbol{q}_t | \boldsymbol{\theta}') \cdot p(\boldsymbol{\theta}') \mathrm{d}\boldsymbol{\theta}'},
\end{equation} 
where $p(\boldsymbol{q}_t | \boldsymbol{\theta})$ is the likelihood, determined by the model, and $p(\boldsymbol{\theta})$ is the prior, which encodes all previous knowledge of $\boldsymbol{\theta}$. The normalization at the denominator is the model's evidence.  

Even though often parameter inference relies on likelihood optimization, the likelihood is intractable in many cases of practical interest, even for  minimal models, like in the case we are discussing here. In fact, the likelihood for $\boldsymbol{q}_t$ is a marginalization (projection) of the full likelihood, $p(\boldsymbol{q}_t | \boldsymbol{\theta}) = \int \mathcal{D}\boldsymbol{x}_t  p(\boldsymbol{q}_t, \boldsymbol{x}_t | \boldsymbol{\theta})$ , which is a path-integral over all possible hidden trajectories $\boldsymbol{x}_t$. It is, in general, analytically intractable and computationally costly. This significantly hinders conventional approaches that require many repeated evaluations of the likelihood or its gradient. SBI is a powerful way to perform Bayesian inference avoiding the evaluation of intractable likelihoods \cite{cranmer_frontier_2020}. Neural Posterior Estimation (NPE), a specific SBI algorithm, aims to directly estimate the posterior from the data \cite{papamakarios_fast_2016,lueckmann_flexible_2017,greenberg_automatic_2019}. 

\subsection{Neural posterior estimation of smFS}

The main ingredients of SBI are an experimental observation, a simulator, and a prior (Fig. \ref{fig: simulation-based inference}). The simulator $\mathcal M (\boldsymbol{\theta})$ is a computer program encoding a parametric model that should explain the observed data. The simulator implicitly encodes the model's likelihood---even if intractable. For any parameter choice $\boldsymbol{\theta}^{(i)}$, the simulator samples the implicit likelihood producing synthetic data $\boldsymbol{q}^{(i)}_t \sim \mathcal M(\boldsymbol{\theta}^{(i)})$ that ideally should reproduce the experimental observation $\boldsymbol{q}^{(\mathrm{obs})}_t$. Drawing $N$ parameter samples from the prior $p(\boldsymbol{\theta})$, the simulator produces a data-set $\mathcal{D} = \{(\boldsymbol{\theta}^{(i)}, \boldsymbol{q}^{(i)}_t\}_{i=1}^N$. In NPE, we use a neural network of parameters $\phi$ to model a conditional density estimator $f_{\phi}(\boldsymbol{\theta}|\boldsymbol{q}_t)$ and train it on $\mathcal{D}$. The trained network is a surrogate of the posterior and allows us to perform inference for any given observation, $p(\boldsymbol{\theta} | \boldsymbol{q}_t = \boldsymbol{q}^{\mathrm{(obs)}}_t) \approx f_{\phi}(\boldsymbol{\theta}|\boldsymbol{q}_t = \boldsymbol{q}^{\mathrm{(obs)}}_t)$, at a negligible computational cost.

We applied NPE on synthetic constant-force smFS experiments to extract quantitative models, and study how well the inference matched the ground truth parameters. In every numerical experiment, we ran Brownian dynamics on $G(q,x)$ with a given set of true parameters $\boldsymbol{\theta}^{(o)}$ to obtain a synthetic observation $\boldsymbol{q}^{\mathrm{(obs)}}_t$. We discarded the corresponding hidden time series $\boldsymbol{x}_t$ and projected all time series $\boldsymbol{q}_t$ on a medium-dimensional feature space (see Methods for more details). We used a uniform prior $p(\boldsymbol{\theta})$ defined in a reasonable range of values. 

NPE extracts hidden parameters from incomplete observations with high accuracy and precision. In the first computational experiment, we trained the posterior on only 600 Brownian simulations (Fig. \ref{fig: double_well_results}A-C). To visualize the inference's quality, we plot the marginal posterior distributions of every single parameter $\theta_i$, having integrated out all the remaining ones, $p(\theta_i | \boldsymbol{q}_t = \boldsymbol{q}^{\mathrm{(obs)}}_t) \equiv \int p(\boldsymbol{\theta} | \boldsymbol{q}_t = \boldsymbol{q}^{\mathrm{(obs)}}_t) \cdot \prod_{j\neq i} \mathrm{d}{\theta_j}$.
%
%\begin{equation}\label{eq: marginals}
%    p(\theta_i | \boldsymbol{q}_t = \boldsymbol{q}^{\mathrm{(obs)}}_t) \equiv \int p(\boldsymbol{\theta} | \boldsymbol{q}_t = \boldsymbol{q}^{\mathrm{(obs)}}_t) \cdot \prod_{j\neq i} \mathrm{d}{\theta_j}.
%\end{equation} 
%
The inference is remarkably good, especially for $D_q/D_x$ and $\Delta G^{\ddagger}$. 
For all three parameters, the posterior's peak is close to the true values. The posterior's spread provides the uncertainty of the inference. While this is reasonably precise for $\Delta G^{\ddagger}$ and $D_q/D_x$, it is not for $\kappa_l$. Reducing the uncertainty requires more simulated data. Training over 6,000 Brownian trajectories led to an exceptional inference (Fig. \ref{fig: double_well_results}D-F). 

Obtaining high-quality inference is computationally efficient. We studied the inference quality as a function of the number of simulations (Fig. \ref{fig: double_well_results}G-I). A few thousand simulations are sufficient to provide good estimates of all three parameters. $D_q/D_x$ is the most accessible parameter to extract, probably because contained in the statistics of local fluctuations. The stiffness $\kappa_l$ is the most challenging parameter to extract, and its uncertainty decreases significantly only after approx. 1000 simulations.
%For $D_q/D_x$, the estimate is unbiased, and the uncertainty decreases progressively with the number of simulations. This is the most accessible parameter to extract, probably because contained in the statistics of local fluctuations. 
%The barrier height is systematically overestimated when less than approx. 100 simulations are used to train the model. The stiffness $\kappa_l$ is the most challenging parameter to extract. It is systematically underestimated with a small number of simulations, and the uncertainty decreases significantly only after approx. 1000 simulations.

\subsection{Amortized inference}

Having trained a neural network to approximate the conditional posterior, we can perform inference for new observations without running any additional simulations. The inference is amortized. Further posterior evaluations only require a forward pass of the trained network, which usually takes milliseconds. 

\begin{figure}[htp]
    \centering
    \includegraphics[width=1.\textwidth]{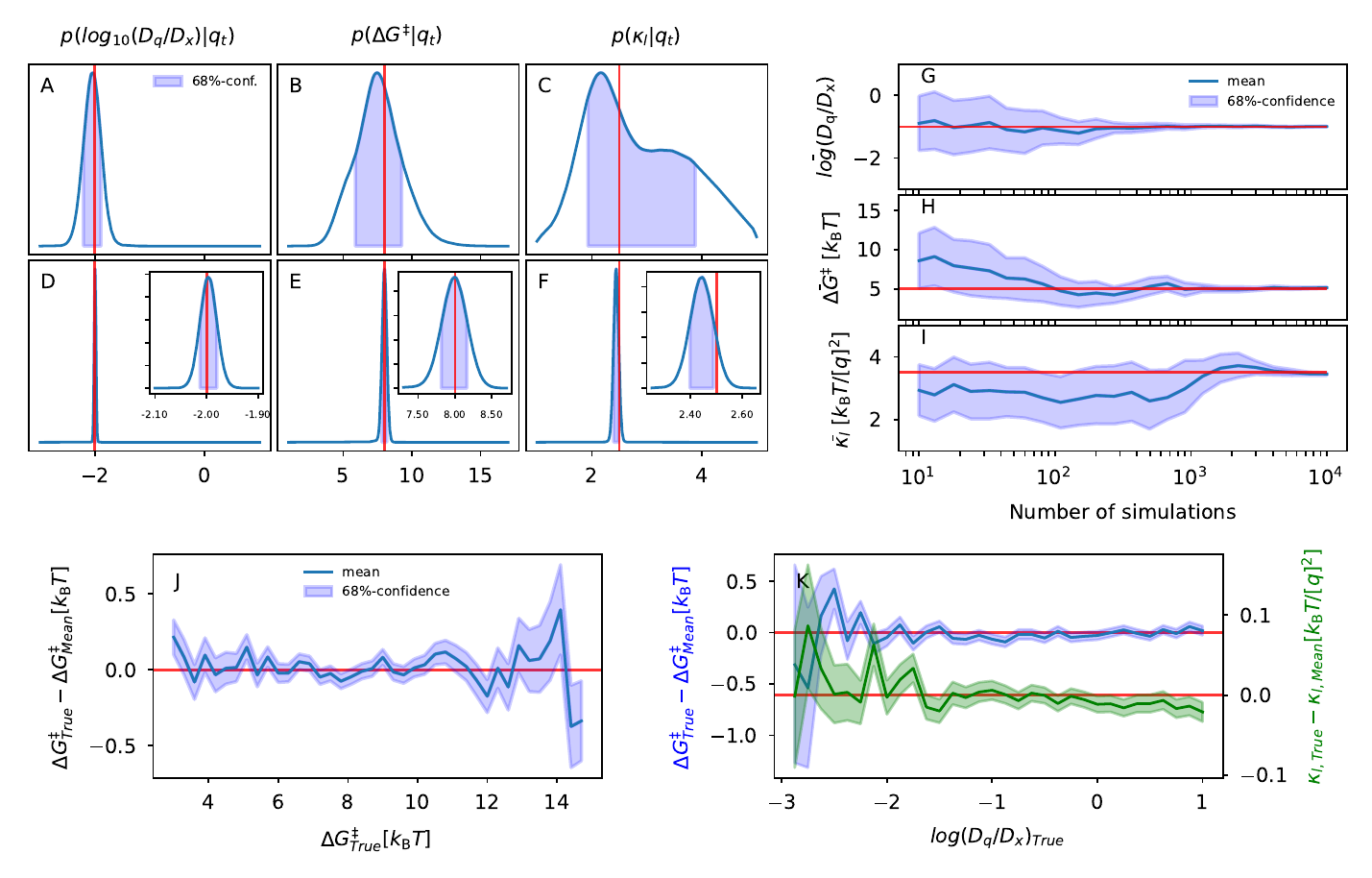}
    \caption[]{\textbf{Neural posterior estimation of smFS at constant force}. (A-F) Posterior marginal distributions as a function of single parameter trained on 600 (A-C) and 6,000 (D-F) simulations. All posteriors are evaluated on the same observation, which was computed with the true parameters indicated by the red vertical lines. The blue shaded area represents $68\%$ of the marginal density, corresponding to a 1$\sigma$ confidence interval. The insets show a zoom-in of the posterior marginals.    
    (G, H, I) Evolution of the posterior marginals as a function of increasing numbers of simulations. For each number of simulations, we trained ten independent posteriors using different training data. The blue line represents the average of the mean of the posteriors, while the blue shaded area is the average of the $1\sigma$ confidence intervals. The observation was generated with the true parameters indicated with the horizontal red line.
    (J) Difference between best estimate and true molecular barrier height as a function of increasing true barrier height. The synthetic observations varied only in the barrier height, while we kept 
    $\log(D_q/D_x) = -1$ and $\kappa_l = 2~k_{\mathrm{B}}T [q]^{-2}$ fixed. We used the mean of the posterior as the best estimate (blue line). (K) Difference between the best estimate and true molecular barrier height as a function of the ratio of diffusion coefficients (blue line) with $\Delta G^{\ddagger} = 7~k_\mathrm{B}T$ kept constant. Difference between the best estimate and true linker stiffness as a function of the ratio of diffusion coefficients (green line) with $\kappa_l = 2~ k_\mathrm{B}T[q]^{-2}$ kept constant. We estimated the $1\sigma$ confidence interval as the $68\%$ of the posterior marginal density.}
    \label{fig: double_well_results}
\end{figure}

We used the posterior trained over 60,000 simulations and systematically investigated the inference's quality. The estimate of the hidden barrier height is excellent for values between 4 and 13 $k_\mathrm{B}T$ (Fig. \ref{fig: double_well_results}J). Lower barriers do not produce clear transitions between the two states. Larger barriers cause poor transition statistics in the training set. Yet, the largest error is only a fraction of $k_\mathrm{B}T$. 
We could obtain excellent estimates of the barrier height and linker stiffness varying $D_q/D_x$ over four orders of magnitude (Fig. \ref{fig: double_well_results}K). The quality degrades for very small values of $D_q$. 
%Fig. \ref{fig: double_well_results}I confirms that $\kappa_l$ is the most demanding parameter to estimate. However, this parameter characterizes the stiffness of the pulling device and can be estimated independently, enabling a more informative prior. 
These data show that SBI allows us to extract accurate diffusive models from synthetic data of smFS experiments over a broad range of parameters.

\subsection{Hidden states and model comparison}

%Bayesian inference  enables model comparison, quantifying how well two alternative models $\mathcal{M}_1$ and $\mathcal{M}_2$ explain the observed data. 
%Usually, this comparison relies on the calculation of the Bayes factor, defined as the ratio of the models' evidences introduced in Eq. \ref{eq: bayes_theorem}. Intractable likelihoods---like in our case---hinder calculating Bayes factors. 
%Here, we performed model comparison by formulating nested models, where  $\mathcal{M}_1$ is a particular case of the more general model $\mathcal{M}_2$. 

An interesting question in the context of smFS is whether we can detect a ``hidden" metastable state. By measuring a single quantity---the observed extensions---we project an inherently high-dimensional dynamical system on a one-dimensional coordinate $q$. If the projection of two meta-stable states leads to similar values of $q$,  we might not resolve one of them. 

\begin{figure}[htp]
    \centering
    \includegraphics[width=0.9\textwidth]{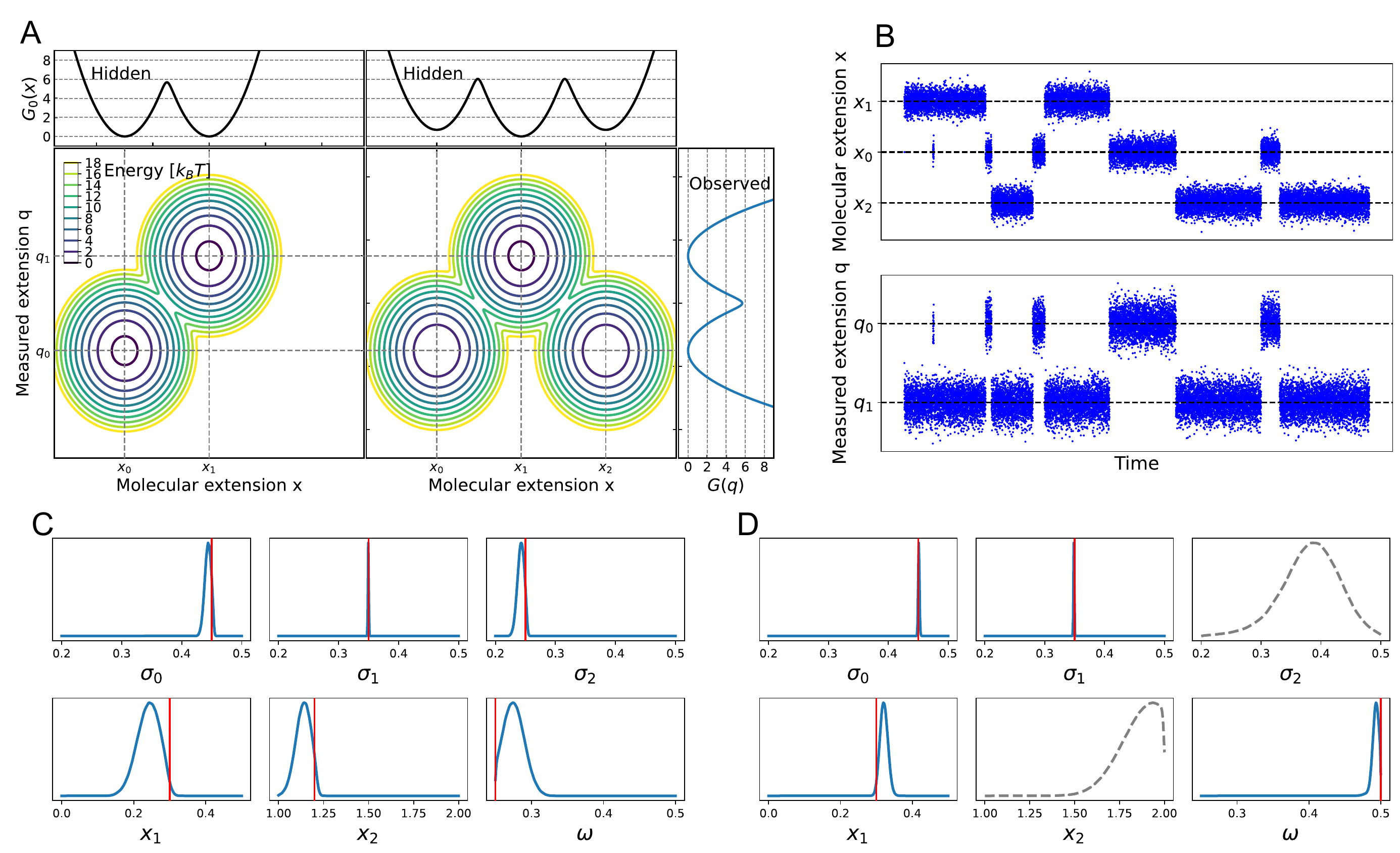}
    \caption[]{\textbf{Model comparison}. (A) We compare the inference of an observation using a two- or three-state models. In the latter, one state is hidden due to the projection on $q$. The molecular free energy profile $G(x)$ shows two and three states, while the potential of mean force is the same and only exhibits two states. (B) Brownian simulations in the three-state model show jumps between 3 states in $x$, but only two states in $q$. (C, D) Posterior marginals obtained by NPE trained on simulations of a nested model. In general, the model contains 3 Gaussian states. The parameter $\omega$ determines their relative weight. For $\omega=0.5$, the model contains only two Gaussian states. (C) Marginals of the posterior $p(\boldsymbol{\theta}|\boldsymbol{q}_t=\boldsymbol{q}^{(3)}_t)$, evaluated on a synthetic observation produced by a three-state model. $p(\omega)$ favours a three-state model to explain this observation. (D) Marginals of the posterior $p(\boldsymbol{\theta}|\boldsymbol{q}_t=\boldsymbol{q}^{(2)}_t)$, evaluated on a synthetic observation produced by a two-state model. $p(\sigma_2)$ and $p(x_2)$ are dashed because not necessary.}
    \label{fig: nested}
\end{figure}

To investigate this problem in a simplified setting, we considered two alternative models, $\mathcal{M}_1$ and $\mathcal{M}_2$. $\mathcal{M}_1$  is defined by free energy surface $G_1(q, x)$, defined by two states (Fig. \ref{fig: nested}A). The first state is centered at $(q_0, x_0)$, while the second is at $(q_1, x_1)$. This model is very similar to the harmonic-linker model studied in the previous section, with the linker implicitly modeled by the relative position and the width of the two states. We then considered a model $\mathcal{M}_2$ with the surface $G_2(q, x)$, defined by three states. The two bottom states are centered at the coordinates $(q_0, x_0)$ and $(q_0, x_2)$; and the top state is centered at $(q_1, x_1)$  (Fig. \ref{fig: nested}A). In the two bottom states, the molecular extension takes different values $x_0$ and $x_2$, but both are projected on the same value $q_0$ (Fig. \ref{fig: nested}B). The time series of $\boldsymbol{q}_t$ produced by both models $\mathcal{M}_1$ and $\mathcal{M}_2$ describe a hopping process between only two states. Can an inference tell us whether a two-state model is enough to explain the observed data or if we need a three-state one?  

The posterior trained on a nested model enables us to choose between models of different complexity. We trained a conditional posterior $p(\boldsymbol{\theta} | \boldsymbol{q}_t )$ using Brownian simulations on a free energy surface $G(q,x)$ defined by a linear combination of three Gaussian distributions (Eq. \ref{eq:gaussian_mix}). The model's parameters are    
%
%\begin{equation}\label{eq: 3_state_density}
%    \rho_{\mathrm{eq}}(x, q | \boldsymbol{\theta}) = \omega \Pi_0(x, q | x_0, \sigma_0 ) + \omega \Pi_1(x, q | x_1, \sigma_1 ) + (1 - 2 \omega ) \Pi_2(x, q | x_2, \sigma_2 ),
%\end{equation}
%
%where $\Pi_i(x, q | x_i, \sigma_i )$ for $i=0,1,2$ are Gaussian distributions with parameters
%
$\{x_i, \sigma_i\}$ for $i=1,2$, which describe location and width, and the mixing coefficient $\omega$, which determines the relative weight between the states. 
%For state $i=0$, only the width $\sigma_0$ could be changed to avoid degenerate solutions. 
This model contains, in general, three states, like $\mathcal{M}_2$, but for $\omega=1/2$ it reduces to the simpler two-state model $\mathcal{M}_1$. We compared the posterior of a synthetic observed time series produced from the three-state model, $p(\boldsymbol{\theta} | \boldsymbol{q}_t = \boldsymbol{q}_t^{(3)})$ (Fig. \ref{fig: nested}C), and one produced with a two-state model, $p(\boldsymbol{\theta} | \boldsymbol{q}_t = \boldsymbol{q}_t^{(2)} )$ (Fig.  \ref{fig: nested}D). In the first case, the marginal posteriors  peak around the true values for all parameters of the three Gaussians. Also, the marginal of $\omega$ indicates that we need a three-state model to make an inference on the ${\boldsymbol{q}_t^{(3)}}$ observation. In the second case, instead, the posterior is sharply peaked around $\omega=1/2$, indicating that a two-state model is sufficient.  

\subsection{Robustness to model misspecification}

SBI performs an excellent inference if the synthetic observed data are produced by the same model that we encoded in the simulator. But what happens if this is not true? In reality, our model will only be an approximation of the process that produced the experimental observation.  

\begin{figure}[htp]
    \centering
    \includegraphics[width=\textwidth]{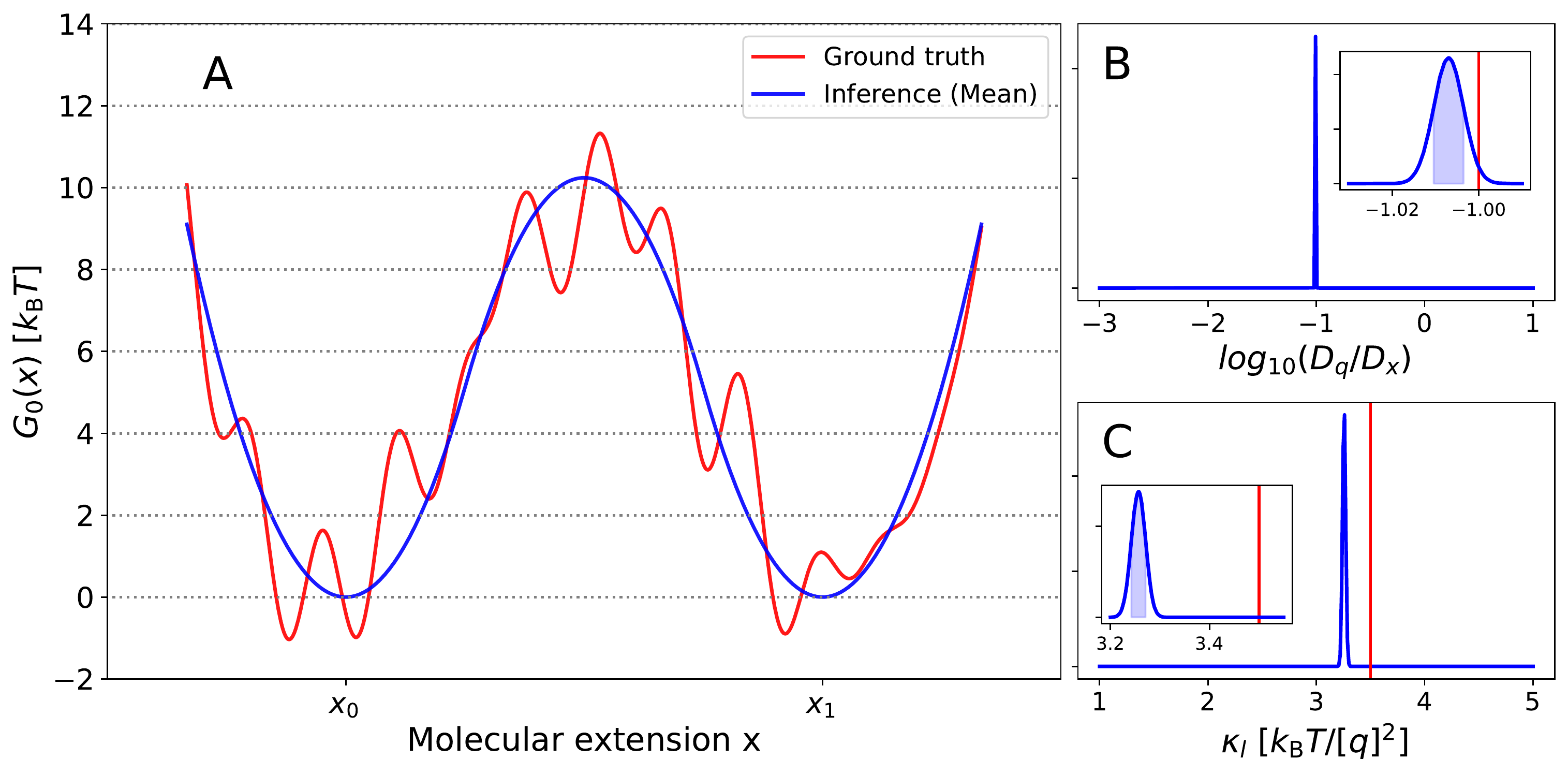}
    \caption[Posterior marginals double well system]{\textbf{Model misspecification}. (A) Best inference  assuming an idealized symmetric double well (blue line), evaluated on a synthetic observation produced by a rough molecular profile $G_0^{r}(x)$ (red line). (B) and (C) Marginal posterior distributions  (blue lines) for the ratio of diffusion coefficients and linker stiffness, respectively. Red lines indicate the parameter true values, while blue shaded areas indicate a 68\% confidence interval.}
    \label{fig: perturbation}
\end{figure}

\begin{figure}[htp]
    \centering
    \includegraphics[width=\textwidth]{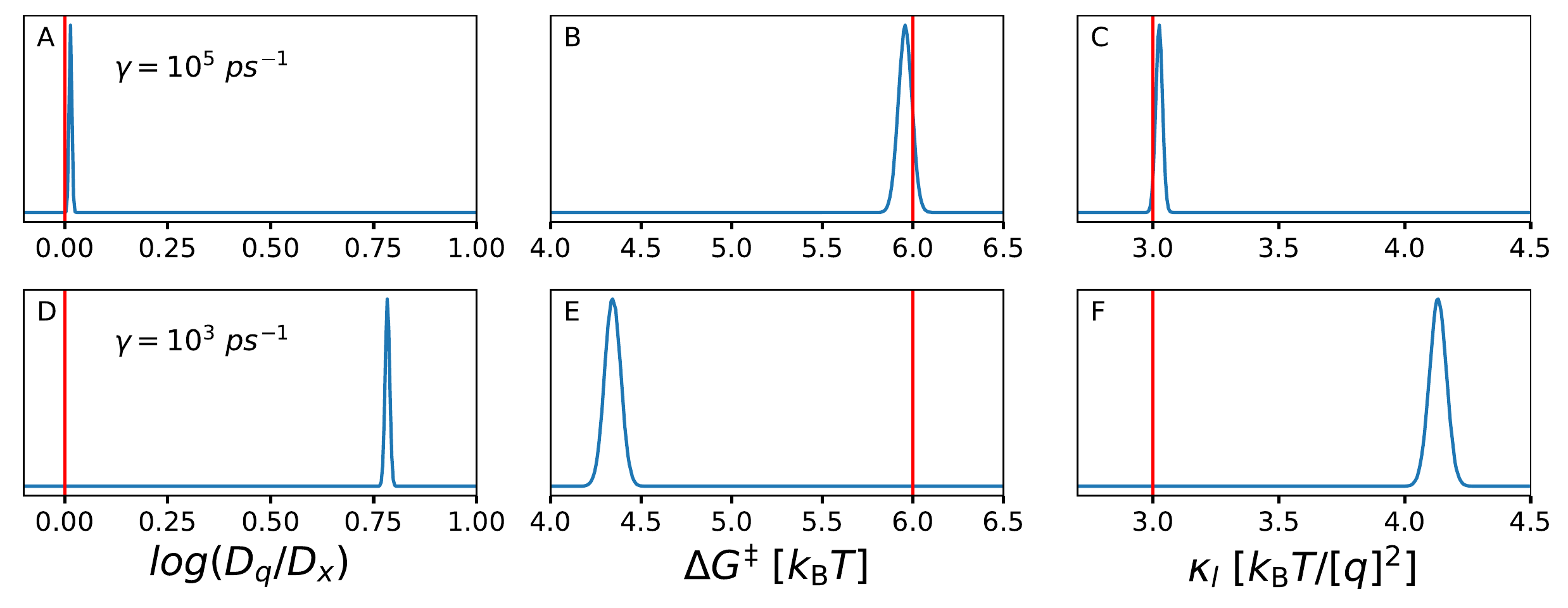}
    \caption[]{\textbf{Inference with dynamics misspecification} (A-C) Posterior marginals trained on over-damped (Brownian) simulations evaluated on an observation produced with a under-damped Langevin integrator at high friction $\gamma = 10^5~\mathrm{ps^{-1}}$ and (D-F) low friction $\gamma = 10^3~\mathrm{ps^{-1}}$. The red line indicates the true values of all parameters.}
    \label{fig: posterior_langevin}
\end{figure}

A moderate model mismatch slightly degrades the estimate's accuracy. If a rough two-state curve $G^{\mathrm{rough}}_{0}(x)$ generated $\boldsymbol{q}_t^{\mathrm{obs}}$, performing inference with the posterior trained assuming a smooth symmetric double-well model will return the best fit to the true curve (Fig. \ref{fig: perturbation}A). The estimated $D_q/D_x$ is very accurate since this quantity depends only on local fluctuations (Fig. \ref{fig: perturbation}B). The estimated $\kappa_l$ is close to the true value (Fig. \ref{fig: perturbation}C). 
However, The posterior severely underestimate the uncertainties. It is too narrow and does not include the true value of $\kappa_l$.  We also considered misspecification of the system's dynamics. Making an inference on synthetic observations produced with inertial dynamics (under-damped Langevin), while assuming a  diffusive one, leads to good results in the limit of high friction (SI Fig.  \ref{fig: posterior_langevin}A-C), but breaks down for low friction (SI Fig.  \ref{fig: posterior_langevin}D-F). 

\begin{figure}[htp]
    \centering
    \includegraphics[width=\textwidth]{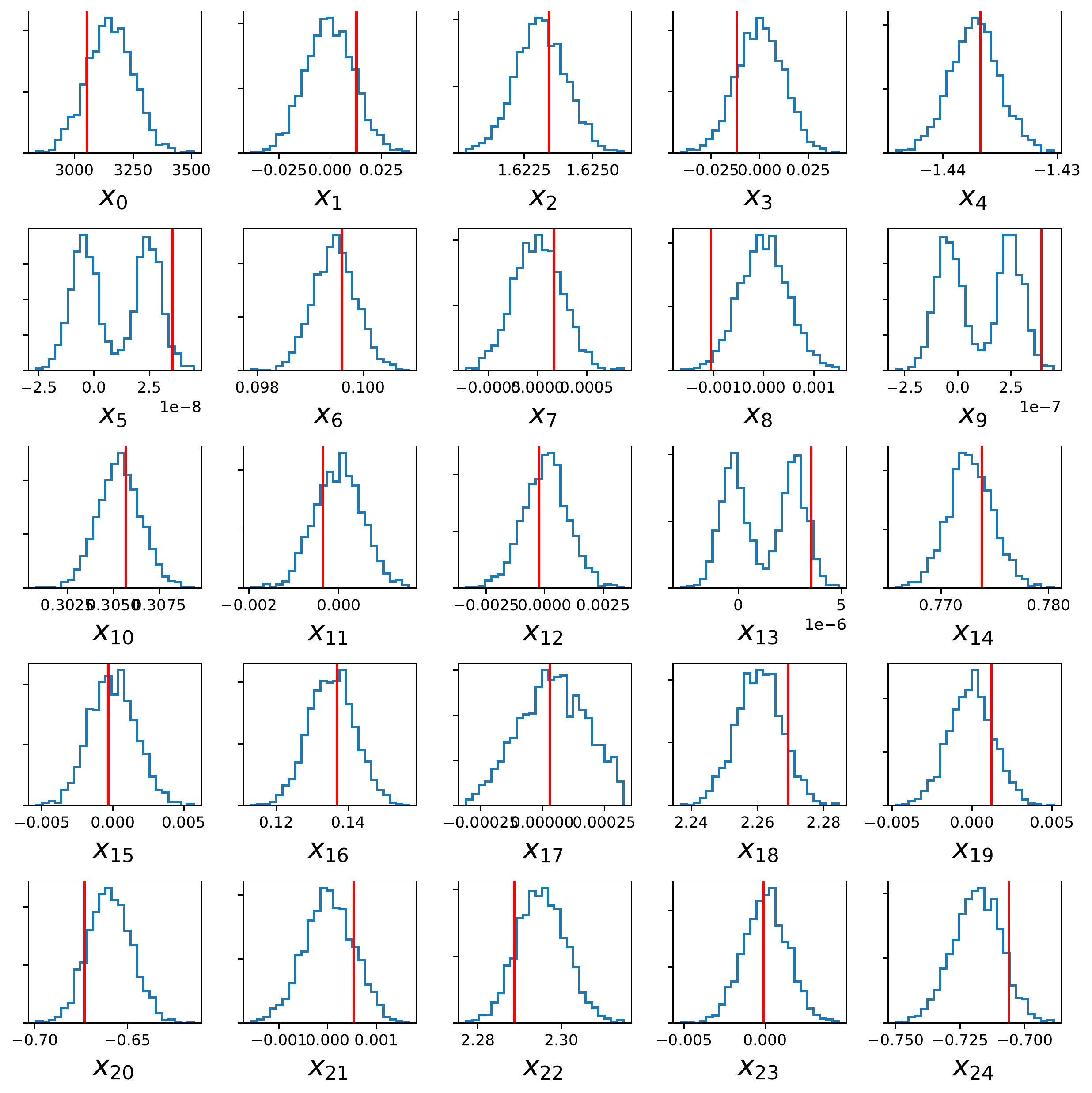}
    \caption[]{\textbf{Posterior predictive check in the absence of a molecular free energy model misspecification}. We inferred the parameters of a synthetic observation produced by Brownian simulator on a smooth double-well potential, and used these parameters to run 2,000 simulations with the same simulator. Each panel shows the histograms of all the 25 features that we used to obtain a medium-dimensional projection of the simulated time series. The red lines indicate the true values, i.e., the values of the features corresponding to the initial synthetic observation. The support of the simulated data contains all true values. }
    \label{fig: SI_ppc_double_well_true}
\end{figure}

\begin{figure}[htp]
    \centering
    \includegraphics[width=\textwidth]{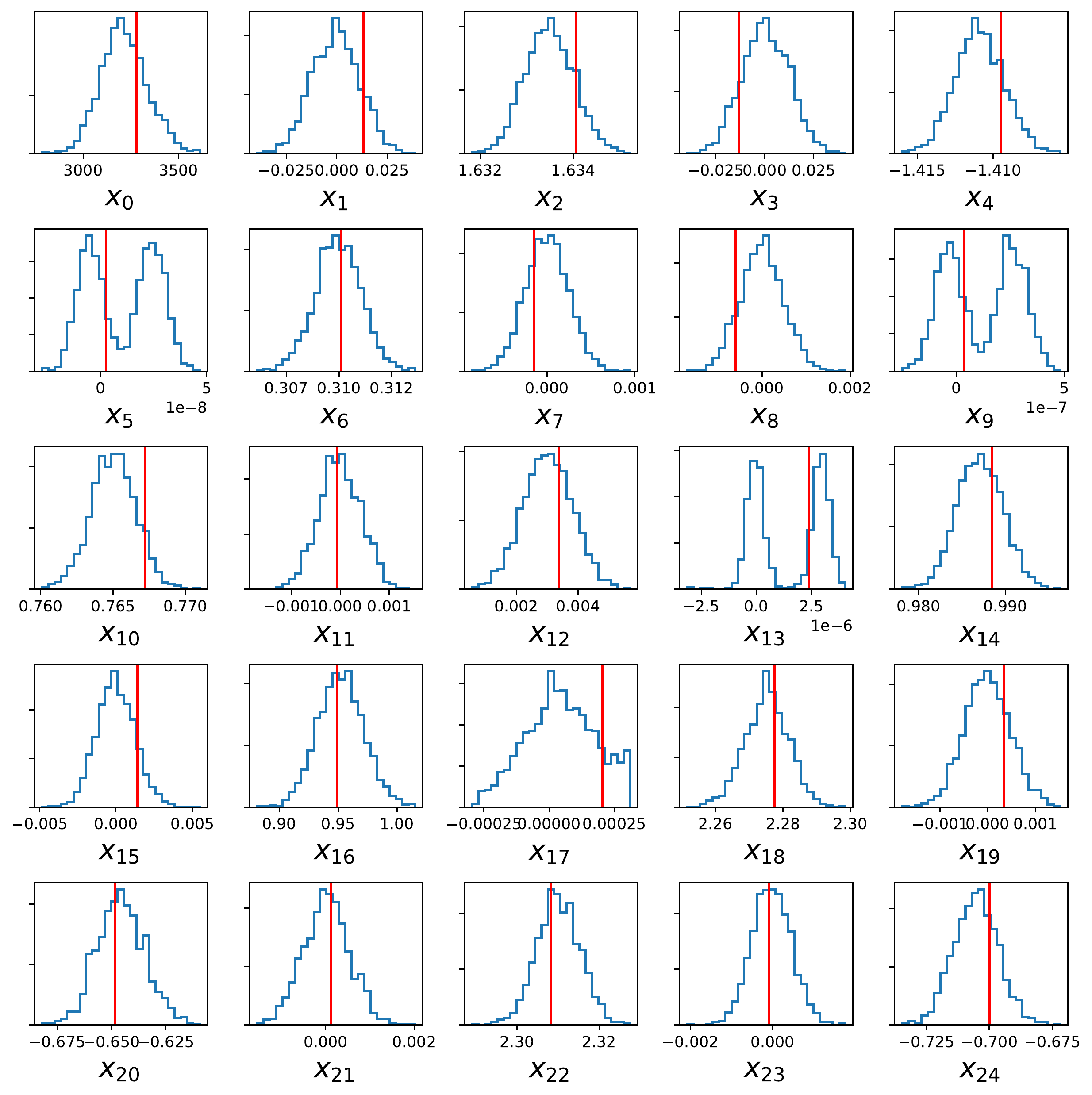}
    \caption[]{\textbf{Posterior predictive check in the absence of dynamics misspecification.} We inferred the parameters of a synthetic observation produced by under-damped Langevin simulator with high friction ($\gamma = 10^5~\mathrm{ps^{-1}}$) on a symmetric double-well, and used these parameters to run 2,000 simulations using a Brownian dynamics integrator on the same energy surface. Each panel shows the histograms of all the 25 features that we used to obtain a medium-dimensional projection of the simulated time series. The red lines indicate the true values, i.e., the values of the features corresponding to the initial synthetic observation. The support of the simulated data contains all true values.}
    \label{fig: SI_ppc_langevin_high_friction}
\end{figure}

\begin{figure}[htp]
    \centering
    \includegraphics[width=\textwidth]{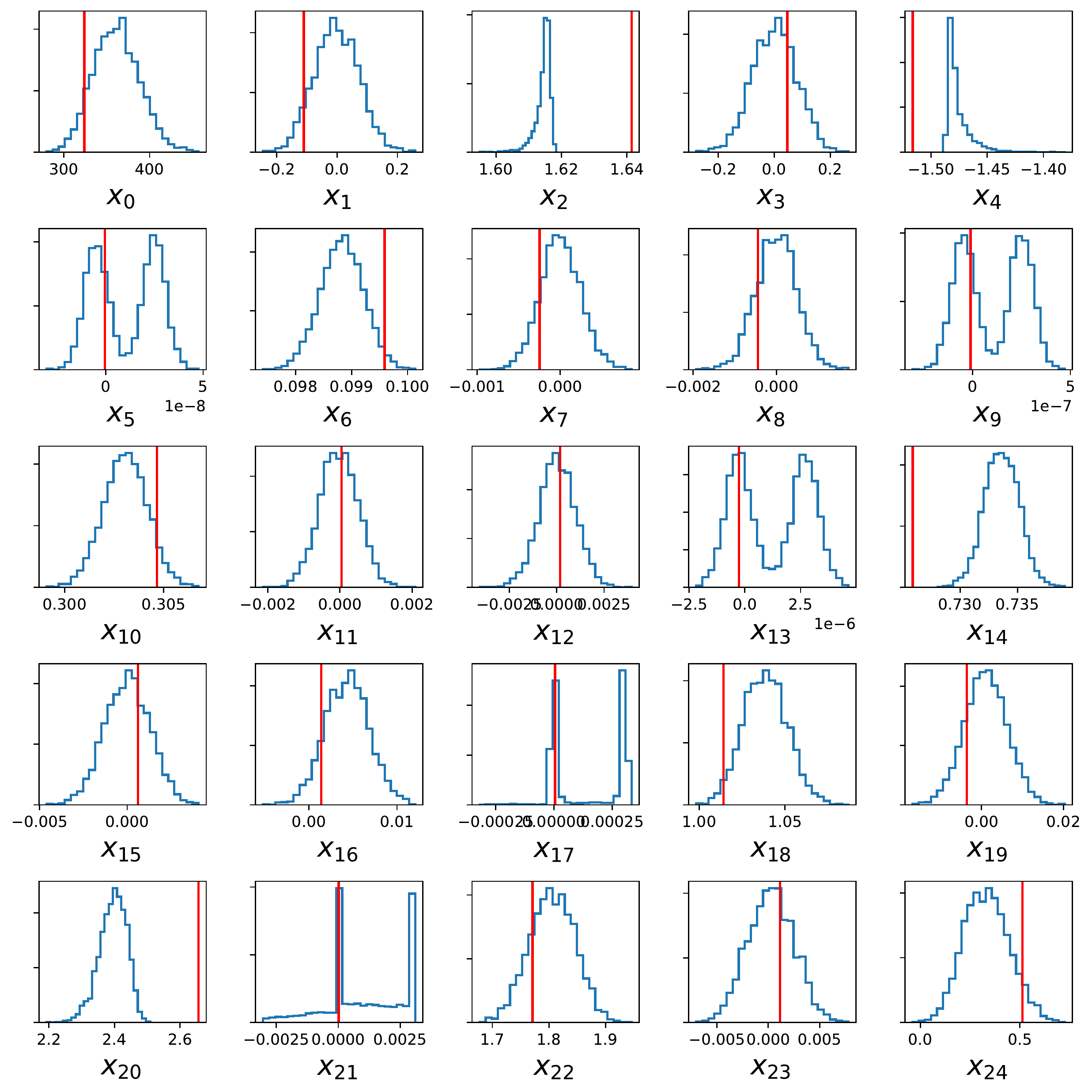}
    \caption[]{\textbf{Posterior predictive check with model misspecification of molecular free energy profile}. We inferred the parameters of a synthetic observation produced by Brownian simulator on a rough double-well potential, and used these parameters to run 2,000 simulations on the smooth symmetric double-well. Each panel shows the histograms of all the 25 features that we used to obtain a medium-dimensional projection of the simulated time series. The red lines indicate the true values, i.e., the values of the features corresponding to the initial synthetic observation. The support of the simulated data does not contain all true values. }
    \label{fig: SI_ppc_double_well_false}
\end{figure}

\begin{figure}[htp]
    \centering
    \includegraphics[width=\textwidth]{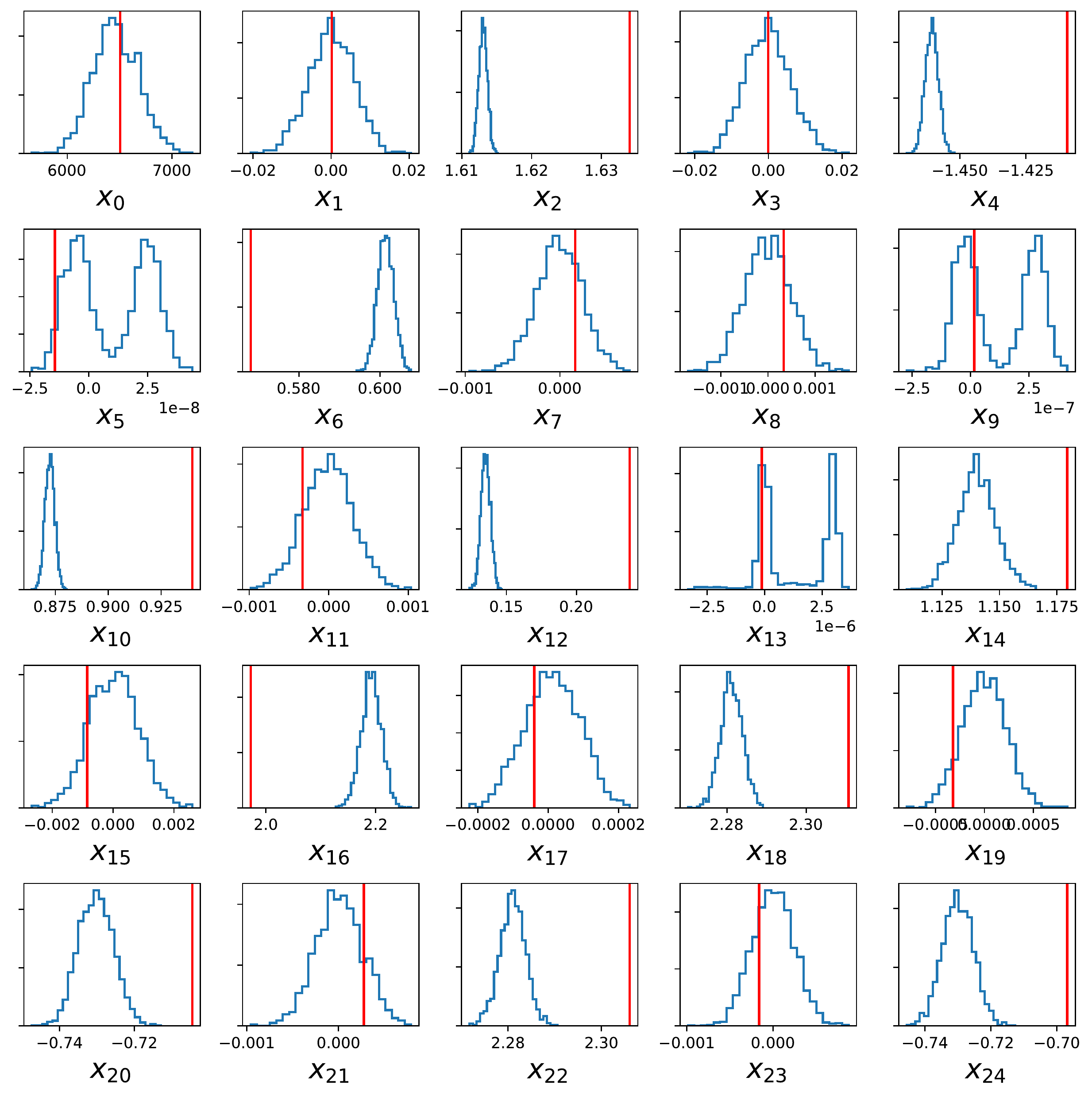}
    \caption[]{\textbf{Posterior predictive check with dynamics misspecification.} We inferred the parameters of a synthetic observation produced by under-damped Langevin simulator with low friction ($\gamma = 10^3~\mathrm{ps^{-1}}$) on a symmetric double-well, and used these parameters to run 2,000 simulations using a Brownian dynamics integrator on the same energy surface. Each panel shows the histograms of all the 25 features that we used to obtain a medium-dimensional projection of the simulated time series. The red lines indicate the true values, i.e., the values of the features corresponding to the initial synthetic observation. The support of the simulated data does not contain all true values.}
    \label{fig: SI_ppc_langevin_low_friction}
\end{figure}

SBI provides tools to diagnose model misspecification. The posterior predictive check reveals whether the experimental observation is "unusual" compared to the simulated data (SI Figs. \ref{fig: SI_ppc_double_well_true} and \ref{fig: SI_ppc_langevin_high_friction}). If so, the inference should not be trusted. Synthetic observations produced with the rough double well or an inertial dynamics are atypical compared to simulations performed in the smooth symmetric double well with Brownian dynamics (SI Figs. \ref{fig: SI_ppc_double_well_false} and \ref{fig: SI_ppc_langevin_low_friction}).

\subsection{Inference of complex free energy landscapes}

Having proven the potential of the SBI approach, we aimed to making inferences of more realistic free energy profiles. Describing folding and other conformational rearrangements generally requires profiles presenting several long-lived intermediates and barriers of different heights. We considered a new class of models for the $G_0(x)$ using polynomial splines. These allow for  greater flexibility than the models considered so far, albeit at the cost of increasing the parameter space, requiring eleven parameters corresponding to the height of the nodes. The total dimension of $\boldsymbol{\theta}$ is now 13.  

\begin{figure}[htp]
    \centering
    \includegraphics[width=\textwidth]{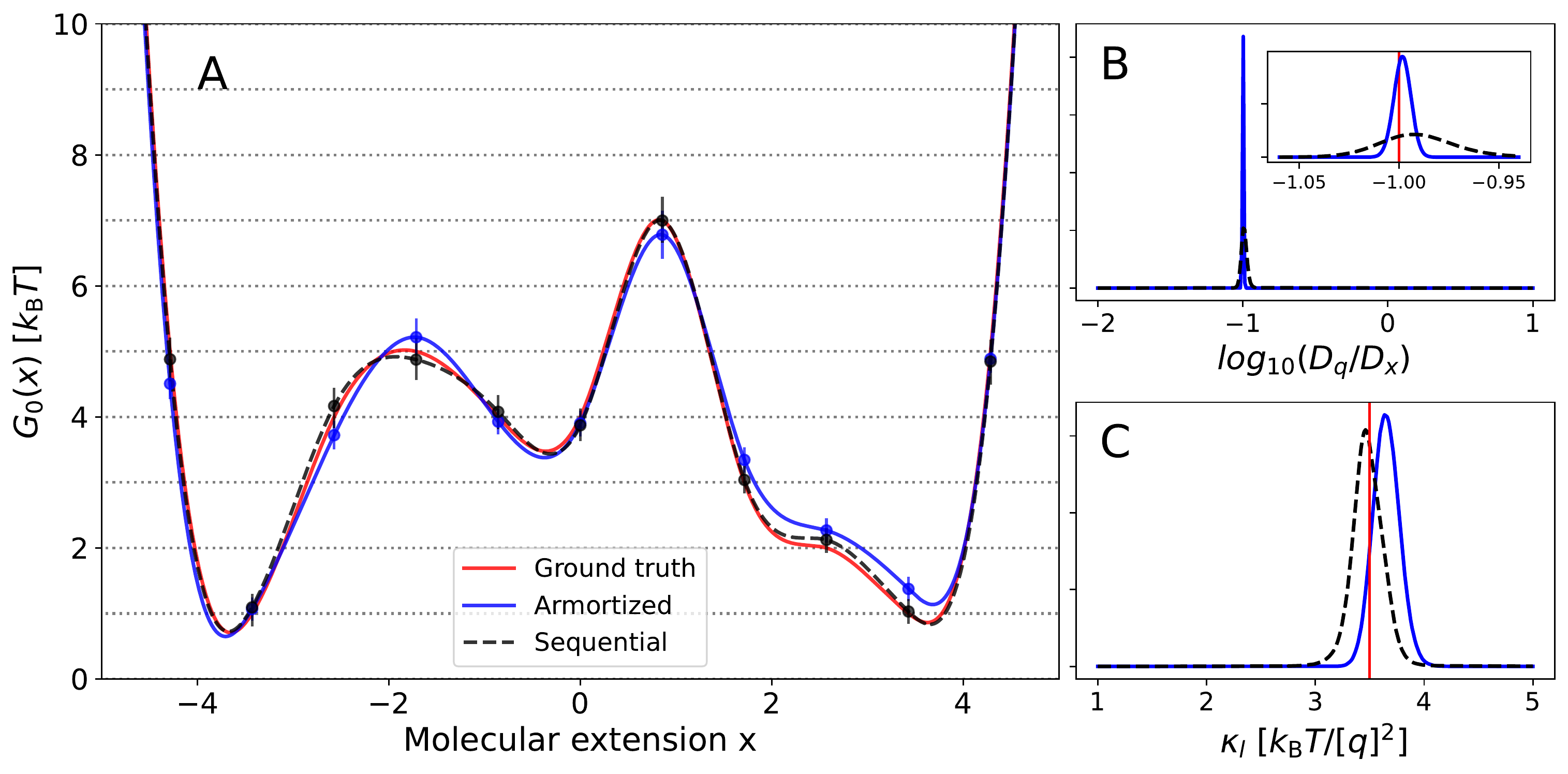}
    \caption[]{\textbf{Amortized and sequential SBI of complex free energy profiles}. (A) Best inference of a complex hidden molecular free energy profile $G_0(x)$. The true hidden profile is indicated in red, while the best estimate obtained from an amortized and sequential inference in blue and dashed black, respectively. Best estimate and error bars are the mean and the $68\%$ marginal density of the posterior, respectively. (B) Posterior marginal as a function of the ratio of diffusion coefficients, and (C) the linker stiffness. The inset in (B) shows a zoom-in.}
    \label{fig: spline_curve}
\end{figure}

\begin{figure}[htp]
    \centering
    \includegraphics[width=\textwidth]{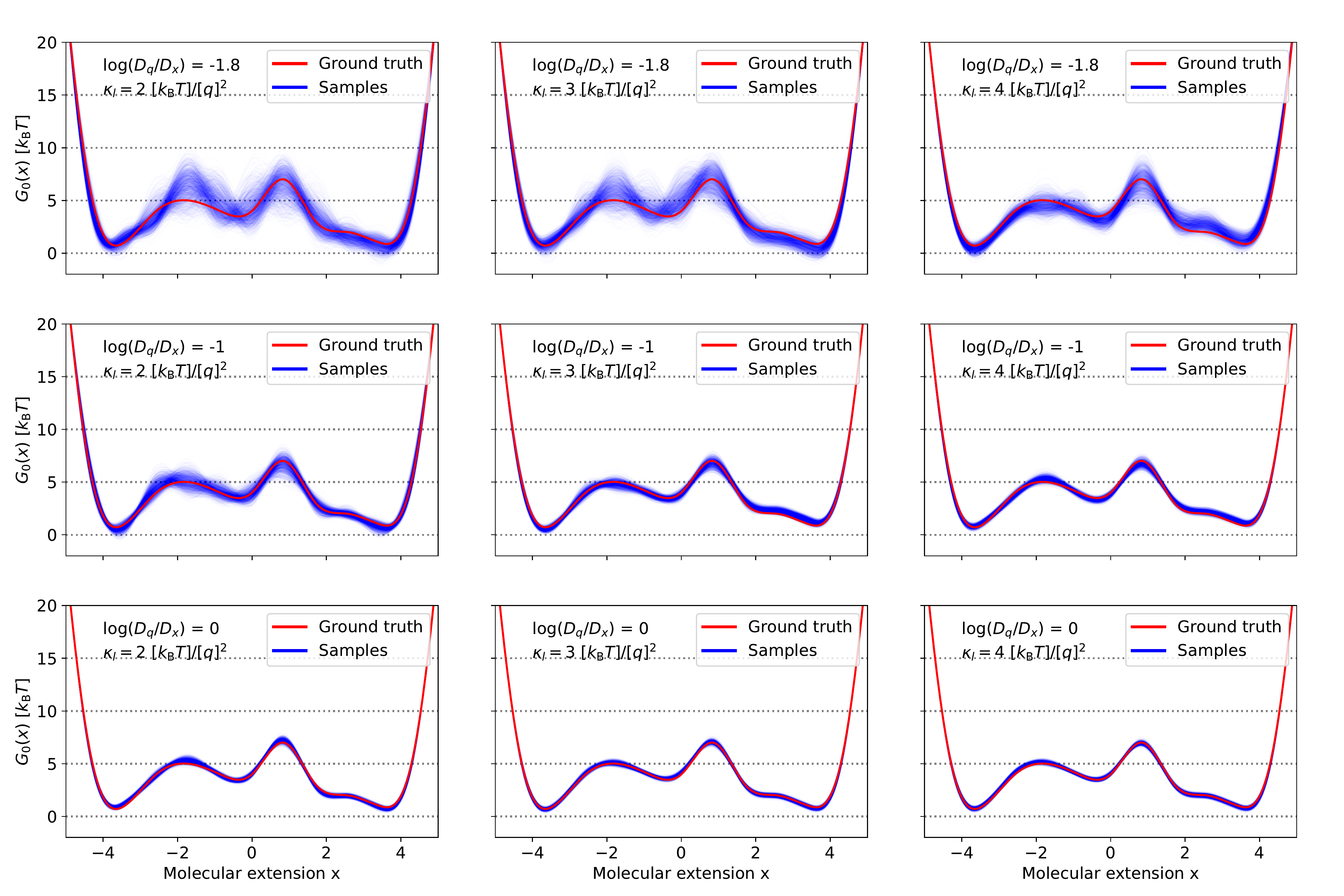}
    \caption[]{\textbf{Systematic amortized inference of complex free energy landscape}. We performed amortized inference of 9 synthetic observations obtained with the same molecular free energy curve $G_0(x)$ (in red), varying the true values of $D_q/D_x$ and $\kappa_l$. For each observation, we generated 500 posterior samples (blue lines). In each of the three columns, we fixed the ratio of diffusion coefficients $\log_{10}D_q/D_x$ to the values $-1.8, -1, 0$, respectively. In each of the three rows the linker stiffness $\kappa_l$ is constant with values values $2,\,3,\,4~[k_{\mathrm{B}}T]/[q]^{-2}$, respectively.}
    \label{fig: SI_complex_spline_systematic}
\end{figure}

Despite the increased complexity, SBI successfully extracts complex hidden molecular free energy profiles. We trained a posterior on simulations performed with the flexible spline model of $G_0(x)$, and extracted profiles with an average error smaller than a $k_{\mathrm{B}}T$ (Fig. \ref{fig: spline_curve}A). The estimate is also very good for the diffusion coefficients of the linker stiffness (Fig. \ref{fig: spline_curve}B and C). The inference is amortized, and allows us to show without further simulations that we can extract $G_0(x)$ over a broad range of parameters (SI Fig. \ref{fig: SI_complex_spline_systematic}). Notably, the spline model does not require defining the number and location of states and barriers. Both are automatically extracted from the observed time series. We obtained this result training on 1.5 million simulations. Whereas for our minimal model this required only a few days of simulations, the same might not be possible for more complex simulators.

\begin{figure}[htp]
    \centering
    \includegraphics[width=\textwidth]{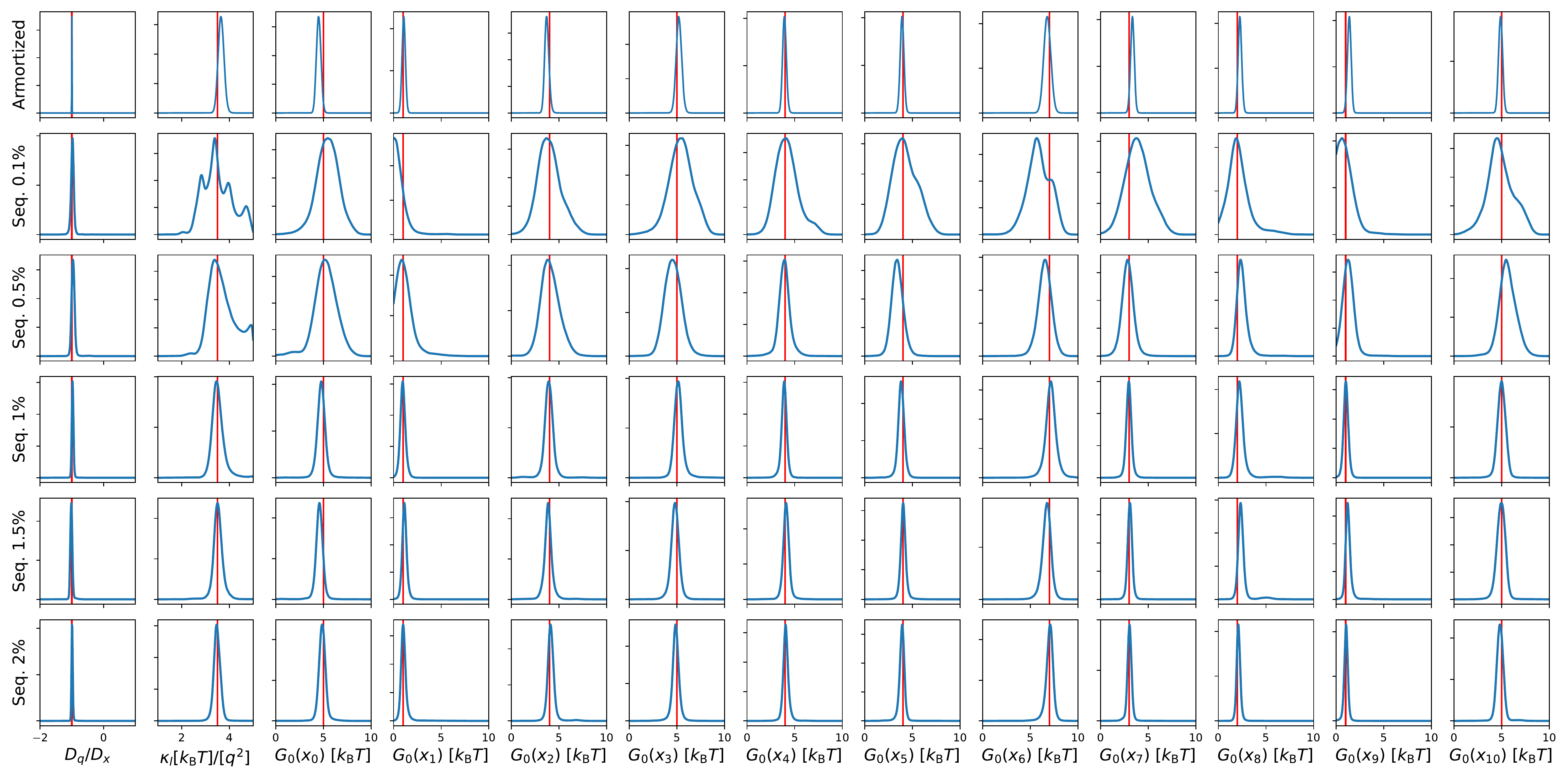}
    \caption[]{\textbf{Sequential posterior inference of complex free energy landscape}. We performed sequential NPE for a single synthetic observation. Each column shows the posterior marginal for all the parameters $\boldsymbol{\theta}$ of this model. The first row shows the marginals obtained for this observations with the amortized posterior trained on 1.5 million simulations. Successive rows show the same marginals of the sequentially trained posterior. Starting from a uniform prior, each row shows the posterior trained on a cumulative amount of simulations expressed as a percentage of the 1.5 million used to train the amortized posterior. Red lines indicate the true values.}
    \label{fig: SI_complex_spline_sequential}
\end{figure}

Sequential SBI is a powerful alternative for computationally expensive simulators, or if we are interested in making inference on a single observation. In this approach, we iterate between running small batches of simulations, training a posterior, and using this posterior as a proposal distribution to initiate new simulations. This is a form of active learning: the algorithm autonomously learns where it should run simulations in the parameter space. In 20 iterations, the sequential approach provided an excellent inference (Fig. \ref{fig: spline_curve}), and used only 30,000 simulations---2\% of the simulations used for the amortized posterior (SI Fig. \ref{fig: SI_complex_spline_sequential}).

\section{Discussion}\label{section: discussion}

Despite their great success, the challenge of extracting quantitative models from partial observations hampers the full potential of smFS experiments. This challenge is a fundamental inverse problem. We lose information by projecting a high-dimensional system on a single quantity. Additionally, we measure this quantity via the mediation of an ever-present experimental apparatus that further distort the measurement. 

In this paper, we showcased how machine-learning-empowered SBI is a general, conceptually simple, and powerful technique for addressing these challenges for smFS experiments at constant force. Using synthetic data, we could extract compact interpretable models of increasing complexity that accurately described hidden physical processes over a broad range of parameters.

Model misspecification remains an outstanding problem for every practical inference. One of the clear advantages of SBI is that it provides a self-contained quality control. Posterior predictive checks can reveal that the model encoded in the simulator is unsuitable for reproducing the observed data and that the inference should not be trusted. 
%Bayesian priors and how they affect the inference quality is an extensively debated topic. If a parameter's true value lies outside the prior's range, it will not come out from the inference. On the other hand, priors are a very transparent way of stating and using previous knowledge and assumptions.   

The main advantage of our approach is that the inference is amortized. The heavy computational part---running the simulator with many different choices of parameters---is necessary to train the posterior but has to be done only once for any given choice of model and priors. Once trained, any new inference requires only plugging the observation in the trained posterior and performing a forward-pass of the underlying neural network. Sequential inference offers a powerful alternative in situations where amortized inference is not feasible or desirable. 

As increasingly more challenging smFS experiments are established, approaches like the one we explored here, together with non-parametric Bayesian techniques \cite{bryan_iv_diffraction-limited_2022}, will become more and more crucial. Identifying mechanistic heterogeneity requires monitoring subtle differences in the transition paths, i.e., how molecules re-organize between alternative meta-stable states. Transition paths are particularly affected by kinetic artifacts from measuring devices \cite{Cossio2018}. With more and more available computational power, it will soon be possible to use quasi-atomistic molecular dynamics with SBI. The simulator will then explicitly map trajectories containing molecular structures to measured extensions, keeping into account the specific position of the linkers, beads, and their physics. Moreover, our SBI formalism could be generalized to account for incomplete observations from multiple rebinding events \cite{bullerjahn_reversible_2022}, force-rupture \cite{Hummer2010}, or other types of single-molecule experiments.

The main challenge of applying inference schemes to actual experimental data is to model the noise correctly. Whereas many approaches are often limited to highly idealized noise models---e.g., Gaussian distributed---real noise is generally very complex. A simulation-based approach like ours can consider any noise model that can be encoded in a simulator. This includes not only known functions that would frustrate analytical treatments but also data-driven models of noise obtained, for instance, using machine learning approaches.

The combination of amortized and sequential inference enables the establishment of so-called foundational models. These would consist of amortized posteriors trained on complex, realistic simulations of biophysical experiments, which might require months to simulate and train. Once trained, they can be made available to the community and serve as the proposal distribution of an inexpensive sequential approach to fine-tune the posterior to specific experiments or observations.

Intractable likelihoods are very common for many important problems and hinder analytical investigations.  
On the other hand, generating synthetic data with high-fidelity simulators is very often straightforward. By integrating physics-based parametric models with machine learning density estimate, SBI enables accurate Bayesian inference for models with an intractable likelihood. In this way, it enables us to consider more complex and realistic models that would be otherwise ruled out due to their mathematical intractability, with great potential for applications in biophysics.

\section{Methods}

\subsection{Details of the theoretical smFS model}
The harmonic-linker model introduced by Hummer and Szabo is a well-established model of smFS experiments \cite{Hummer2010,Cossio2015,covino_molecular_2019}. In this model, a two-dimensional free energy surface describes the combined system of molecule and apparatus. The molecular (hidden) extension $x$ is defined as a distance between two amino acids in the protein to which the linkers are attached. A smFS experiment returns the measured extension $q$, which reports the distance between the molecular linkers connecting the molecule and the experimental apparatus. 
The free energy surface $G(q, x)$ is:
\begin{equation}\label{eq: free_energy}
    G(q, x) = G_0(x) + \frac{\kappa_l}{2}(x - q)^2~.
\end{equation}
The first term $G_0(x)$ describes the molecular free energy profile and implicitly includes the constant pulling force of the apparatus applied in smFS experiments at constant-tension. We considered several models for $G_0$, as explained in the next section. The second term describes the spring-like coupling of the molecule to the apparatus by the linker. The parameter $\kappa_l$ describes the stiffens of the linker and $x-q$ is the linker extension.
We assume that the diffusion is position independent and anisotropic, $D_x \neq D_q$. While $D_x$ is an intrinsic molecular property, $D_q$ is mainly governed by the diffusion coefficient of the mesoscopic pulling device.

We obtained trajectories describing  the time evolution the system from an initial position by simulating Brownian dynamics on the free energy surface $G(x, q)$ (Eq. \ref{eq:brown_dyn}). In the simulations, the system will spend most of the time in one of the metastable states while observing fast transitions between them (Fig \ref{fig: SMFE} B). To mimic the situation of a smFS, we kept only $\boldsymbol{q}_t$.

\subsection{Models of the molecular free energy $G_0(x)$}\label{section: methods_free_energy}

We provide the details of the various models we used to describe the  one-dimensional molecular free energy profile $G_0(x)$.

\subsubsection{Symmetric-double well}

In the simplest case, the molecular free energy profile consists of a symmetric bi-stable double well
\begin{equation}\label{eq:matched_parabola1}
    G_0(x) = \Delta G^{\ddagger}\cdot f(x/x^{\ddagger})~,
\end{equation}
where
\begin{equation}\label{eq:matched_parabola2}
    f(x) = 
    \begin{cases}
    - 2x^2,\ for\ 0\leq |x| \leq 1/2\\
    2(|x| - 1)^2,\ for\ 1/2 < |x|,\\
    \end{cases}
\end{equation}
with $\Delta G^{\ddagger}$ is the energy barrier between the two minima at $x=0$. The two meta-stable states are positioned at $x = \pm x^{\ddagger}$ and represent, for instance, the folded and unfolded states of the protein. For all simulations we set $x^{\ddagger} = 1.5\,[q]$. For this model, the parameters that enter the prior and posterior are $\boldsymbol{\theta} = \{ \Delta G^{\ddagger}, D_q/D_x, \kappa_l \}$.

\subsubsection{Two- and three-well states surfaces}

To follow a more general approach, we modelled the full two-dimensional free energy surface $G(q, x)$ using the negative logarithm of a linear combination of multiple two-dimensional Gaussian functions. Therefore, we did not have to explicitly define the potential of the linker molecule, which is implicitly encoded in the relative configuration of the Gaussians. We defined:
\begin{equation}\label{eq:gaussian_mix}
     G(q, x) = -k_{\mathrm{B}}T \ln \left\{ \sum_{i=0}^{K-1} \frac{\omega_i}{2\pi (\sigma_{q,i} \sigma_{x,i})^{\frac{1}{2}}} \exp\left(-\frac{(x-x_i)^2}{2\sigma_{x,i}^2} - \frac{(q-q_i)^2}{2\sigma_{q,i}^2}\right)\right\},
\end{equation}
where $q_i$ and $x_i$ are the positions of the minima of the different states, and $\sigma_{q,i}$, $\sigma_{x,i}$ their widths, along $q$ and $x$, respectively. The $\omega_i$ are the weights of the different states, with $\sum_i \omega_i = 1$. 

To construct the nested model, we used three states. The states where positioned along $q$ at $q_0 = q_2 = -0.75$ and $q_1 = 0.75$. Thus, two of the states $(i = 0,2)$ overlapped along the measured extension. We inferred the x-position of the second and third states $(i =  1,2)$, while the first state was kept fixed at $x_0 = -1.5$. We set the weight for each state such that the projected free energy $G(q)$ is the same symmetric double-well. The weight of the second state was set to $\omega_1 = 0.5$, while the weight of the first and last state are set to $\omega_0 = \omega$ and $\omega_2 = 0.5 - \omega$. By changing the parameter $\omega$ the weight between the first and last states is changed, while $G(q)$ does not change. 

\subsubsection{Rough double well}
To investigate the effect of moderate model mismatch, we produced synthetic experimental time series $\boldsymbol{q}_t$ generated from a rough (noisy) version of the symmetric two well potential introduced in the previous section: $G^{\mathrm{rough}}_{0}(x) = G_0(x) + \eta(x)$. 
The perturbation function $\eta(x)$ is a sum of different sinus functions, with different amplitudes $a_i$, frequencies $b_i$, and phase shifts $c_i$, i.e., $\eta(x) = A \sum_{i=0}^{N-1} a_i \sin(b_i x + c_i)$. We constructed random realizations of this rough potential by drawing the parameters  from uniform distributions. The factor $A$ is a scaling constant controlling the amount of noise added to the molecular free energy profile, and, therefore, the deviation from the idealized profile $G_0(x)$.
We set A was to $0.7$ for all simulations. 

\subsubsection{Flexible spline profile}

We used a cubic spline interpolation from the GNU Scientific Library to build a flexible model of the molecular free energy profile $G_0(x)$. We selected fifteen points $\{(x_i,G_0(x_i))\}$, equally spaced out along the x-axis, and connected them pairwise with a third degree polynomial: $G_0(x) = a_i\cdot x^3 + b_i\cdot x^2 + c_i\cdot x + d_i\ \forall\  x \in [x_i, x_{i+1}]$. The first two and last two nodes had fixed values to avoid the system escaping the energy wells. The first and last node had a fixed values of $G_0(x_0) = G_0(x_{14}) = 70 ~k_\mathrm{B}T$, the second and second last $G_0(x_1) = G_0(x_{13}) = 30~k_\mathrm{B}T$. 

The values of $G_0(x_i)$ for the inner eleven spline nodes $i\in \{2, \dots, 12\}$ specify the details of free energy profile. Every new simulation propagates the system on a different spline $G_0(x)$. Therefore, for this system, the parameters that enter the prior and posterior are $\boldsymbol{\theta} = \{ G_0(x_2), \dots, G_0(x_{12}), D_q/D_x, \kappa_l \}$. After training, we sampled the posterior and aligned the free energy profiles, which are defined up to an additive constant. 

\subsection{Details of the simulator}\label{section: methods_brownian_dynamics}

The simulator integrated the equations of motion according to Brownian dynamics (over-damped Langevin) on a free energy surface $G(q,x)$. We used the the Euler-Maruyama integration scheme
\begin{align}\label{eq:brown_dyn}
    q(t + \Delta t) = q(t) -\beta\partial_q G(q, x)\cdot D_q \Delta t + \sqrt{2 D_q \Delta t}\cdot R_q(t)\\
x(t + \Delta t) = x(t) -\beta\partial_x G(q, x)\cdot D_x \Delta t + \sqrt{2 D_x \Delta t}\cdot R_x(t)~,
\end{align}
where $D_x$ and $D_q$ are the diffusion coefficients along the $q$ and $x$-axis, respectively, and $R_q(t)$ and $R_x(t)$ are uncorrelated Gaussian random numbers with zero mean and a unit variance.  We set the integration time step in all simulations to $D_x \Delta t = 5\cdot10^{-4}$. 
We decimated the raw trajectories to get time series reproducing synthetic experimental data for the measured and molecular extensions, $\boldsymbol{q}_t = \{q_{t{\Delta \tau}}\}_{t=1}^{M}$ and $\boldsymbol{x}_t=\{x_{t{\Delta \tau}}\}_{t=1}^{M}$, respectively, saving $M$ time frames every $\Delta \tau$. For the symmetric double well and the Langevin models we saved $M = 10^8$ frames, saving every $\Delta \tau = 100$; for the nested model $M = 2\cdot 10 ^7$ frames, saving every $\Delta \tau = 50$; and for the cubic spline model $M = 10^6$, saving every $\Delta \tau = 100$.

For the under-damped Langevin simulations, we used the Langevin integrator in OpenMM \cite{eastman2017openmm}. We set the temperature to $500~K$, the mass to $10^{-3}$ atomic units, and the timestep to $5\cdot10^{-4}~\mathrm{ps}$.

\subsection{Time series featurization}\label{section: methods_summary_stats}

Time series are structured (very) high-dimensional data, which cannot be directly used to perform neural density estimation. We must therefore project the original data $\boldsymbol{q}_t$ on a medium-dimensional set of features $\boldsymbol{y} = \boldsymbol{y}(\boldsymbol{q}_t)$. We can either use summary statistics, which might be already available in a given scientific domain, or use an additional neural network that extracts features from the data, e.g., an encoder. Here, we chose to use summary statistics. 

For the double-well models, we described each time series $\boldsymbol{q}_t$ with 25 features $y_i$. We used the number of observed transitions between metastable states per unit time---an estimate of the microscopic rates---the first four statistical moments of the distribution of observed positions $\rho(\{q_i\})$, and of the distribution of displacements $\rho(\{\Delta q_k\})$, with $\Delta q_k = q_{i + k} - q_i$ calculated at five different lag-times  $k = [1, 10, 100, 10000, 100000]$. We estimated the number of transitions based on changes in the running mean $\overline{q}_i = 1/w\sum_{j=-w/2}^{w/2}q_{i+j}$. The parameter $w$ determines the window size and affects the estimate of the number of jumps. However, the final inference does not strongly depend on the estimated rate. 

For the complex spline model, we used the transition matrices $T_{ij}(\Delta \tau)$ for different lag times $\Delta \tau$ as summary statistics. To compute the transitions matrix  $T_{ij}(\Delta \tau)$ the trajectory $\boldsymbol{q}_t$ was binned in 20 equally spaced bins. For each lag time, we populated the transition matrix counting the transitions between bins $i$ and $j$. The matrix was normalized to one along the columns. We computed the transition matrix for the lag times $\Delta \tau = [1, 10, 100, 1000, 10000, 100000]$. We did not use the rate as a feature for simulations obtained on the complex spline molecular free energy profile.

\subsection{Details of simulation-based inference}

We used SBI to perform Bayesian inference with an intractable likelihood \cite{cranmer_frontier_2020}. In particular, we used NPE \cite{tejero-cantero_sbi_2020}, where we approximate the posterior $p(\boldsymbol{\theta} | \boldsymbol{y}(\boldsymbol{q}_t))$ from simulated data with a neural network-based conditional density estimators $f_{\phi}(\boldsymbol{\theta} | \boldsymbol{y}(\boldsymbol{q}_t))$. The neural network model can vary depending on the specific problem. We used mixture density networks and normalizing flows. 

Relatively simple posterior distributions can be approximated with  mixture density networks (MDN) \cite{bishop1994mixture}. MDN are a general framework to approximate conditional densities with a superposition of $K$ Gaussians: 
\begin{equation}\label{eq: MDN}
    f_{\phi}(\boldsymbol{\theta}|\boldsymbol{y}) = \sum_{k=1}^K \alpha_k \mathcal{N}(\boldsymbol{\theta}|\boldsymbol{m}_k, \boldsymbol{S}_k),
\end{equation}
where the means $\boldsymbol{m}_k$, covariance matrices $\boldsymbol{S}_k$ and mixing coefficients $\{\alpha_k\}$ are all non-linear functions of the observation $\boldsymbol{y}$, approximated by a neural network of parameters $\phi$. To train the network, we maximised the average log probability 
$\frac{1}{N}\sum_{i=1}^{N} \log f_{\phi}(\boldsymbol{\theta^{(i)}}| \boldsymbol{y}^{(i)}(\boldsymbol{q}_t^{(i)}))$
w.r.t. $\phi$ on the training set $\mathcal{D} = \{(\boldsymbol{\theta}^{(i)}, \boldsymbol{y}^{(i)}(\boldsymbol{q}_t^{(i)})\}_{i=1}^N$. 
%which is equivalent to maximising the likelihood $\prod_{i=1}^{N} f_{\phi}(\boldsymbol{\theta^{(i)}}| \boldsymbol{y}^{(i)})$.

For more complex models, we instead used normalizing flows, an alternative approach to estimating conditional densities that offer more flexibility \cite{lueckmann_flexible_2017}\cite{greenberg_automatic_2019}. A normalizing flow is a series of invertible mappings to transform a simple base distribution into a complex target distribution \cite{kobyzev_normalizing_2021}. 
%A simple base distribution---like a multidimensional conditional Gaussian $\pi(\boldsymbol{\theta}|\boldsymbol{y})$---is transformed by the invertible function $f_{\phi}^{-1}(\boldsymbol{\theta},\boldsymbol{y})$ into the target distribution, the posterior $p(\boldsymbol{\theta}|\boldsymbol{y})$. The transformation is defined by the following equation:
%
%\begin{align}\label{eq: flow}
%    p(\boldsymbol{\theta}|\boldsymbol{y}) = \pi(f_{\phi}^{-1}(\boldsymbol{\theta},\boldsymbol{y})| \boldsymbol{y})\cdot \det \left| \left(\frac{df_{\phi}^{-1}(\boldsymbol{\theta},\boldsymbol{y})}{\boldsymbol{y}}\right)\right|.
%\end{align}
%
%The function $f(\boldsymbol{\theta}, \boldsymbol{y})$ is often chosen to be a composition of invertible neural networks with parameters $\phi$. The normalizing flow is trained by maximizing the log-likelihood of $f_{\phi}$ w.r.t. $\phi$. The choice of $f_{\phi}$ is critical, because it determines the flexibility of the flow, thus the capabilities to model complex distributions. 
We used the neural spline flow, which uses cubic splines parameterized by neural networks to model $f_{\phi}$ \cite{durkan2019neural}. We used the implementation of both MDN and neural spline flows available at the SBI package \cite{tejero-cantero_sbi_2020}.

We trained both the mixture density network and the neural spline flow using the Adam optimizer. We adjusted the specific training settings and hyper-parameters for each problem separately (see \ref{section: priors_hyperparam}). We terminated the training after the validation loss did not improve for a given number of epochs.

\subsection{Priors and Hyperparameters}\label{section: priors_hyperparam}
\subsubsection{Symmetric double-well}\label{section: double_well_appendix}
\textbf{Prior}. We chose uniform distributions covering reasonable values based on previous publications \cite{Cossio2015,covino_molecular_2019}: $\log_{10}(D_q/D_x) \sim \mathcal{U}(-3, 1),\ 
    \Delta G^\ddagger \sim \mathcal{U}(3~k_\mathrm{b}T, 17~k_\mathrm{b}T),\ 
    \kappa_l \sim \mathcal{U}(1~k_\mathrm{b}T/[q]^{-2}, 5~k_\mathrm{b}T/[q]^{-2})$.\\
\textbf{Hyperparameters}. We used an MDN as the density estimator, with $K=50$ and a feed-forward neural network with three layers. Each layer had 80 hidden nodes (sometimes called features in the SBI literature) and used the ReLU activation function. The output from the third layer yielded the parameters of the Gaussians. We kept $25\%$ of the simulation data for the validation. The batch size was the default value of 50. We stopped training after the validation loss did not improve for over 40 epochs. The training took 267 seconds on one Intel Core i9-12900K processor. 

\subsubsection{Nested model}\label{section: nested_appendix}
\textbf{Prior}. We chose the priors to ensure that the configuration of the three Gaussian distributions would reproduce the essential features of a smFS experiment: $ \sigma_1 \sim \mathcal{U}(0.2, 0.5), \sigma_2 \sim \mathcal{U}(0.2, 0.5),\sigma_3 \sim \mathcal{U}(0.2, 0.5), x_1\sim \mathcal{U}(0.5, 1), x_2 \sim \mathcal{U}(1, 2), \omega \sim \mathcal{U}(1/4, 1/2)$. \\
\textbf{Hyperparameters}.
We used an MDN as the density estimator with $K=50$, with the same neural network topology as in \ref{section: nested_appendix}. We only increased the number of hidden features per layer to 150. We kept $15\%$ of the simulated data for validation. The batch size was set to 500. We stopped the training after the validation loss did not improve for over 20 epochs. The training took 446 seconds on one Intel Core i9-12900K processor. 

\subsubsection{Cubic spline}\label{section: spline_appendix}
\textbf{Prior}. For the diffusion coefficients and linkers we used the similar priors as for the symmetric double-well $\log_{10}(D_q/D_x) \sim \mathcal{U}(-2, 1),\ \kappa_l \sim \mathcal{U}(1~k_\mathrm{b}T/[q]^{-2}, 5~k_\mathrm{b}T/[q]^{-2})$. The prior for the internal spline values were instead $G_{0, i} \sim \mathcal{U}(0~k_{\mathrm{B}}T, 10~k_{\mathrm{B}}T)\ \forall i \in \{2,\dots,12\}$. \\
\textbf{Hyperparameters}. We used a neural spline flow as the density estimator as implemented in the SBI-toolkit \cite{tejero-cantero_sbi_2020}. We used 5 transformations with 100 hidden features. We augmented the neural spline flow with an convolutional layer with 6 input channels and 6 output channels. The kernel had a size of 6x6 and a stride of 2. The convolutional layer used an ReLu activation function. We kept $15\%$ from the simulation data for validation. The batch size was set to 1500. The training stopped after the validation loss did not improve for more than 20 epochs. The training took 27778 seconds on one Xeon Skylake Gold 6148 Processor. 

For the sequential approximation of the posterior, we iteratively ran new simulations with parameters from a prior or the previous posterior. The new simulations were added to the training set and the approximate posterior was further trained. We used the SNPE-C version of the sequential posterior estimation from the SBI-Toolbox\cite{tejero-cantero_sbi_2020}. In total, we performed 20 rounds of approximation each adding 1500 new simulations to the data set. We used the same hyperparameters for the training and the density estimator as for the amortized case. The cumulative training time for all 20 sequential runs was 24619 seconds on one Intel Core i9-12900K processor.

\subsection{Code}
We generated, analysed, and visualized the data with custom code based on Numpy \cite{harris2020array}, Scipy \cite{virtanen_scipy_2020}, Numba \cite{lam2015numba}, Cython \cite{behnel2010cython}, Pytorch \cite{paszke2019pytorch} and Matplotlib \cite{hunter2007matplotlib}.
We performed the spline interpolation using the implementation of the GNU Scientific library \cite{galassi2002gnu}.
We used the simulation-based inference algorithm NPE and the implementation of the mixture-density network and the neural spline flow from the SBI-Toolkit \cite{tejero-cantero_sbi_2020}. 

\subsection{Code and Data Availability}
Code and data are available on a public repository \cite{dingeldeindata}.

\section{Acknowledgements}
We thank Drs. Attila Szabo and J{\"u}rgen K{\"o}finger for useful discussions and feedback. L.D. and R.C. acknowledge the support of the Frankfurt Institute of Advanced Studies, the LOEWE Center for Multiscale Modelling in Life Sciences of the state of Hesse, and the CRC 1507: Membrane-associated Protein Assemblies, Machineries, and Supercomplexes. R.C. acknowledges the support of the International Max Planck Research School on Cellular Biophysics. P.C. acknowledges the Simons Foundation. We thank the Center for Scientific Computing of the Goethe University and the Juelich Supercomputing Centre for computational resources and support. R.C. dedicates this paper in loving memory of Piero Angela, chief Italian science journalist, who, in more than 50 years of incessant activity, has narrated the beauty of science to many generations. 

\bibliography{Dingeldein2022.bib}

\newpage

\section{Supplementary material}

\end{document}